\documentclass[aps,pre,preprint,showpacs,10pt]{revtex4}
\usepackage{amsmath}
\usepackage{graphicx}

\begin{document}

\title{Transport properties of dense dissipative hard-sphere fluids for
arbitrary energy loss models.}
\author{James F. \surname{Lutsko}}
\affiliation{Center for Nonlinear Phenomena and Complex Systems\\
Universit\'{e} Libre de Bruxelles\\
Campus Plaine, CP 231, 1050 Bruxelles, Belgium}
\email{jlutsko@ulb.ac.be}
\date{\today }

\begin{abstract}
The revised Enskog approximation for a fluid of hard spheres which lose
energy upon collision is discussed for the case that the energy is lost from
the normal component of the velocity at collision but is otherwise
arbitrary. Granular fluids with a velocity-dependent coefficient of
restitution are an important special case covered by this model. A normal
solution to the Enskog equation is developed using the Chapman-Enskog
expansion. The lowest order solution describes the general homogeneous
cooling state and a generating function formalism is introduced for the
determination of the distribution function. The first order solution,
evaluated in the lowest Sonine approximation, provides estimates for the
transport coefficients for the Navier-Stokes hydrodynamic description. All
calculations are performed in an arbitrary number of dimensions.
\end{abstract}

\maketitle

\section{Introduction}

Hydrodynamics is important because it provides a connection between
microscopic models of particle interactions and experimentally observable
behavior. Its experimental significance is due to the fact that
hydrodynamics describes the evolution of fundamental quantities, including
local mass, momentum and energy density which are important in a variety of
applications from microfluidics to cosmology. At a microscopic level, these
quantities are ''slow'' variables which evolve on a timescale which is well
separated from faster, microscopic variables so that the effect of the
latter can be adequately encapsulated in the various transport coefficients
appearing in the hydrodynamic description. Recently, much interest has
centered on the hydrodynamics of granular media which are characterized by a
loss of energy during collisions between the grains\cite%
{GranularPhysicsToday}, \cite{GranularGases}, \cite{GranularGasDynamics}.
The simplest microscopic model of granular materials consists of hard-sphere
grains which lose a fixed fraction $\varepsilon $ of the part of the kinetic
energy associated with the longitudinal, or normal, component of the
velocities at the moment of collision while the transverse components of the
velocities of the colliding particles are unchanged. The transport
properties of this model system, which will be referred to as a ''simple
granular gas'' below, have been worked out in some detail\cite%
{DuftyGranularTransport},\cite{HCS_Mix}. However, there is interest in more
complex models in which the fractional energy loss $\varepsilon $ is itself
a function of the normal kinetic energy as such models are apparently more
realistic\cite{Brilliantov},\cite{Poshel},\cite{McNamara}. Furthermore,
there are a wealth of phenomena, such as endo- and exo-thermic chemical
reactions, in which kinetic energy gets converted to some other form and
thus couples, e.g., chemical reactions and hydrodynamics. A good example is
somnoluminescence\cite{SomnoReview},\cite{Bubbles} where classical
hydrodynamics is often used, with some success\cite{Gaspard}, to try to
understand the complex processes taking place under extreme conditions.
Recently, the possible scattering laws for dissipative collisions consistent
with conservation of momentum and angular momentum has been discussed and
formally exact expressions for the balance equations of mass, momentum,
energy and species have been formulated and the Enskog approximation
discussed\cite{LutskoJCP}. The purpose of the present work is to show that
the Chapman-Enskog expansion \cite{McLennan}can be applied to that kinetic
theory for the case of a one-component fluid in $D$ dimensions with an
arbitrary model for the normal energy loss and reduced to a relatively
simple form thereby providing convenient expressions for the transport
coefficients covering this entire class of models. These results will
therefore include as special cases the known results for elastic
hard-spheres in 2 and 3 dimensions\cite{Gass},\cite{McLennan} and for the
simple granular gas in 3 dimensions\cite{DuftyGranularTransport} as well as
such interesting models as those with velocity-dependent coefficient of
restitution\cite{Brilliantov},\cite{Poshel},\cite{McNamara} for which the
Enskog equation has not previously been solved.

The emphasis on instantaneous interactions is due to their unique
properties. Instantaneous hard-core interactions can be described in the
Enskog approximation which is a finite density approximation known to give
reasonable results for moderately dense fluids\cite{McLennan} and which can
even describe hard-core solids\cite{KDEP}. For most other interactions, only
the Boltzmann description, a low density approximation, is available. The
fact that the Enskog kinetic model is applicable to solids means that there
is scope for application to dense granular systems, which might involve
jamming, as well as to extreme conditions such as occur during
somnoluminescence. Thus, despite its artificial nature, the hard-core model
is an important tool in understanding the real world.

In Section II of this paper, the possible scattering laws, exact balance
laws and consequent Enskog equations are reviewed. The general form of the
Chapman-Enskog expansion is also discussed. The zeroth-order Chapman-Enskog
result is just the exact description of a spatially homogeneous system and
is discussed in Section III. For an equilibrium system, this is the Maxwell
distribution but when there is energy loss, the fluid cools and the
resulting homogeneous but non-stationary state is known as the Homogeneous
Cooling State (HCS). Section III formulates the description of the HCS in
terms of a generating function formalism and gives all information needed to
calculate the simplest corrections to the Maxwell distribution for arbitrary
energy-loss models. In Section IV, the Chapman-Enskog expansion is extended
to first order which is sufficient to get the Navier-Stokes transport
properties. The general formalism is illustrated by application to the
simple granular gas in $D$ dimensions. The paper concludes with a discussion
of the approximations made and comparison to more complete calculations.

\section{Theoretical Background}

\subsection{Kinetic theory with energy loss}

We consider a collection of particles having mass $m$ and hard sphere
diameter $\sigma $ which interact via instantaneous collisions. The position
and velocity of the $i$-th particle will be denoted by $\overrightarrow{q}%
_{i}$ and $\overrightarrow{v}_{i}$ respectively and the the combined phase
variable $\left( \overrightarrow{q}_{i},\overrightarrow{v}_{i}\right) $ will
sometimes be denoted as $x_{i}$. The only scattering law allowing for energy
loss that is still consistent with conservation of total momentum and
angular momentum\cite{LutskoJCP} is that two particles having relative
velocity $\overrightarrow{v}_{12}=\overrightarrow{v}_{1}-\overrightarrow{v}%
_{2}$ prior to collision must have relative velocity 
\begin{equation}
\overrightarrow{v}_{12}^{\prime }=\overrightarrow{v}_{12}-\widehat{q}%
_{12}\left( \overrightarrow{v}_{12}\cdot \widehat{q}_{12}+sgn\left( 
\overrightarrow{v}_{12}\cdot \widehat{q}_{12}\right) \sqrt{\left( 
\overrightarrow{v}_{12}\cdot \widehat{q}_{12}\right) ^{2}-\frac{4}{m}\delta E%
}\right)
\end{equation}%
after the collision, where $\widehat{q}_{12}$ is a unit vector pointing from
the center of the first atom to the center of the second atom. The energy
loss is $\delta E$ which we allow in general to be a function of the normal
relative velocity%
\begin{equation}
\delta E=\Delta \left( \overrightarrow{v}_{12}\cdot \widehat{q}\right) .
\end{equation}%
and the center of mass velocity $\overrightarrow{V}_{12}=\overrightarrow{v}%
_{1}+\overrightarrow{v}_{2}$ is unchanged. It is easy to confirm that the
change of energy upon collision is%
\begin{equation}
\frac{1}{2}mv_{1}^{\prime 2}+\frac{1}{2}mv_{2}^{\prime 2}=\frac{1}{2}%
mv_{1}^{2}+\frac{1}{2}mv_{2}^{2}-\delta E.
\end{equation}%
The simple granular fluid model is based on the energy loss function%
\begin{equation}
\Delta \left( x\right) =\left( 1-\alpha ^{2}\right) \frac{m}{4}x^{2}
\end{equation}%
where the constant $\alpha $ is the coefficient of restitution. More complex
models involve a velocity dependent $\alpha $ while other choices of energy
loss function, involving e.g. a thresholds for energy loss, would be
appropriate for chemical reactions. In the following, it is convenient to
introduce the momentum transfer operator $\widehat{b}$ defined for any
function of the velocities $g\left( \overrightarrow{v}_{1},\overrightarrow{v}%
_{2};t\right) $ by%
\begin{equation}
\widehat{b}g\left( \overrightarrow{v}_{1},\overrightarrow{v}_{2};t\right)
=g\left( \overrightarrow{v}_{1}^{\prime },\overrightarrow{v}_{2}^{\prime
};t\right) .
\end{equation}%
In some applications, the energy loss may not occur for all collisions but
rather might be a random occurrence. We will therefore also consider
throughout a somewhat generalized model in which for any particular
collision, the energy loss function is randomly chosen from a set of
possible functions $\left\{ \Delta _{a}\left( x\right) \right\} $ with
probability $K_{a}\left( \widehat{q}_{12}\cdot \overrightarrow{v}%
_{12}\right) $ which may itself, as indicated by the notation, depend on the
dynamic variable $\widehat{q}_{12}\cdot \overrightarrow{v}_{12}$. A simple
case would be the a fixed fraction $K_{1}=1-p$ of collisions are elastic
with energy loss function $\Delta _{1}\left( x\right) =0$ while the
remainder occur with probability $K_{2}=p$ are inelastic with energy loss $%
\Delta _{2}\left( x\right) \neq 0$. In any case, it is assumed that $%
\sum_{a}K_{a}\left( x\right) =1$ for all $x$. The momentum transfer operator
for the type $a$ collisions will be written as $\widehat{b}_{a}$.

The kinetic theory, Liouville equation and the Enskog approximation, for
arbitrary energy loss function has been discussed in ref. \cite{LutskoJCP}.
The one body distribution function $f\left( \overrightarrow{q}_{1},%
\overrightarrow{v}_{1};t\right) $, giving the probability to find a particle
at position $\overrightarrow{q}_{1}$ with velocity $\overrightarrow{v}_{1}$
at time $t$, satisfies, in the Enskog approximation, the kinetic equation%
\begin{equation}
\left( \frac{\partial }{\partial t}+\overrightarrow{v}_{1}\cdot \frac{%
\partial }{\partial \overrightarrow{q}_{1}}\right) f\left( x_{1};t\right) =J%
\left[ f,f\right]
\end{equation}%
where the shorthand notation $x_{1}=\left( \overrightarrow{q}_{1},%
\overrightarrow{v}_{1}\right) $ is used. The collision operator is%
\begin{equation*}
J\left[ f,f\right] =-\int dx_{2}\;\overline{T}_{-}\left( 12\right) \chi
\left( \overrightarrow{q}_{1},\overrightarrow{q}_{2};\left[ n\right] \right)
f\left( \overrightarrow{q}_{1},\overrightarrow{v}_{1};t\right) f\left( 
\overrightarrow{q}_{2},\overrightarrow{v}_{2};t\right)
\end{equation*}%
where $\chi \left( \overrightarrow{q}_{1},\overrightarrow{q}_{2};\left[ n%
\right] \right) $ is the local equilibrium pair distribution function which
is in general a functional of the local density. The binary collision
operator $\overline{T}_{-}\left( 12\right) $ is 
\begin{equation}
\overline{T}_{-}\left( 12\right) =\left[ \sum_{a}J_{a}\left( \overrightarrow{%
v}_{1},\overrightarrow{v}_{2}\right) \left( \widehat{b}_{a}\right)
^{-1}K_{a}\left( \widehat{q}_{12}\cdot \overrightarrow{v}_{12}\right) -1%
\right] \Theta \left( -\overrightarrow{v}_{12}\cdot \overrightarrow{q}%
_{12}\right) \delta \left( q_{12}-\sigma \right) \overrightarrow{v}%
_{12}\cdot \widehat{q}_{12}
\end{equation}%
where $\Theta \left( x\right) $ is the step function, $\left( \widehat{b}%
_{a}\right) ^{-1}$ is the inverse of the momentum exchange operator and $%
J_{a}$ is the Jacobian of the inverse collision%
\begin{equation}
J_{a}\left( \overrightarrow{v}_{1},\overrightarrow{v}_{2}\right) =\left| 
\frac{\partial \left( \widehat{b}_{a}\overrightarrow{v}_{1},\widehat{b}_{a}%
\overrightarrow{v}_{2}\right) }{\partial \left( \overrightarrow{v}_{1},%
\overrightarrow{v}_{2}\right) }\right| ^{-1}.
\end{equation}%
It will not be necessary to work much with this complicated operator as most
calculations can make use of its simpler adjoint $T_{+}(12)$ defined for
arbitrary functions of the phase variables $A\left( x_{1},x_{2}\right) $ and 
$B\left( x_{1},x_{2}\right) $ by%
\begin{equation}
\int dx_{1}dx_{2}\;A\left( x_{1},x_{2}\right) \overline{T}_{-}\left(
12\right) B\left( x_{1},x_{2}\right) =-\int dx_{1}dx_{2}\;B\left(
x_{1},x_{2}\right) T_{+}\left( 12\right) A\left( x_{1},x_{2}\right)
\end{equation}%
so that%
\begin{equation}
T_{+}\left( 12\right) =\Theta \left( -\overrightarrow{v}_{12}\cdot 
\overrightarrow{q}_{12}\right) \delta \left( q_{12}-\sigma \right) \left( -%
\overrightarrow{v}_{12}\cdot \widehat{q}_{12}\right) \left[
\sum_{a}K_{a}\left( \widehat{q}_{12}\cdot \overrightarrow{v}_{12}\right) 
\widehat{b}_{a}-1\right] .
\end{equation}

\subsection{Hydrodynamic fields and balance equations}

The hydrodynamic fields are the local mass density $\rho \left( 
\overrightarrow{r},t\right) $, the local velocity field $\overrightarrow{u}(%
\overrightarrow{r},t)$ and the local temperature field $T\left( 
\overrightarrow{r},t\right) $. They are defined in terms of the distribution
by%
\begin{eqnarray}
\rho \left( \overrightarrow{r},t\right) &=&mn\left( \overrightarrow{r}%
,t\right) =m\int d\overrightarrow{v}_{1}\;f\left( \overrightarrow{r},%
\overrightarrow{v}_{1};t\right)  \label{fields} \\
\rho \left( \overrightarrow{r},t\right) \overrightarrow{u}(\overrightarrow{r}%
,t) &=&m\int d\overrightarrow{v}_{1}\;\overrightarrow{v}_{1}f\left( 
\overrightarrow{r},\overrightarrow{v}_{1};t\right)  \notag \\
\frac{D}{2}n\left( \overrightarrow{r},t\right) k_{B}T\left( \overrightarrow{r%
},t\right) &=&\frac{1}{2}m\int d\overrightarrow{v}_{1}\;V_{1}^{2}f\left( 
\overrightarrow{r},\overrightarrow{v}_{1};t\right)  \notag
\end{eqnarray}%
where $n\left( \overrightarrow{r},t\right) $ is the local number density and 
$D$ is the number of dimensions. In the third equation $\overrightarrow{V}%
_{1}=\overrightarrow{v}_{1}-\overrightarrow{u}(\overrightarrow{r},t)$. Their
time evolution follows from that of the distribution and is given by\cite%
{LutskoJCP}%
\begin{eqnarray}
\frac{\partial }{\partial t}n+\overrightarrow{\nabla }\cdot \left( 
\overrightarrow{u}n\right) &=&0  \label{balance} \\
\frac{\partial }{\partial t}\rho \overrightarrow{u}+\overrightarrow{\nabla }%
\cdot \left( \rho \overrightarrow{u}\overrightarrow{u}\right) +%
\overrightarrow{\nabla }\cdot \overleftrightarrow{P} &=&0  \notag \\
\left( \frac{\partial }{\partial t}+\overrightarrow{u}\cdot \overrightarrow{%
\nabla }\right) T+\frac{2}{Dnk_{B}}\left[ \overleftrightarrow{P}:%
\overrightarrow{\nabla }\overrightarrow{u}+\overrightarrow{\nabla }\cdot 
\overrightarrow{q}\right] &=&\xi  \notag
\end{eqnarray}%
The pressure tensor is the sum of two contributions $\overleftrightarrow{P}=%
\overleftrightarrow{P}^{K}+\overleftrightarrow{P}^{C}$ with the kinetic part
being 
\begin{equation}
\overleftrightarrow{P}^{K}\left( \overrightarrow{r},t\right) =m\int d%
\overrightarrow{v}_{1}\;f_{l}\left( \overrightarrow{r},\overrightarrow{v}%
_{1},t\right) \overrightarrow{V}_{1}\overrightarrow{V}_{1},  \label{flux1}
\end{equation}%
and the collisional contribution being%
\begin{eqnarray}
\overleftrightarrow{P}^{C}\left( \overrightarrow{r},t\right) &=&-\frac{m}{4V}%
\sigma \sum_{a}\int dx_{1}dx_{2}\;\widehat{q}_{12}\widehat{q}_{12}\left( 
\widehat{q}_{12}\cdot \overrightarrow{v}_{12}\right) \delta \left(
q_{12}-\sigma \right) \Theta \left( -\widehat{q}_{12}\cdot \overrightarrow{v}%
_{12}\right)  \label{flux2} \\
&&\times \chi \left( \overrightarrow{q}_{1},\overrightarrow{q}_{2};\left[ n%
\right] \right) f\left( x_{1};t\right) f\left( x_{2};t\right) K_{a}\left( 
\widehat{q}_{12}\cdot \overrightarrow{v}_{12}\right) \int_{0}^{1}dy\;\delta
\left( \overrightarrow{r}-y\overrightarrow{q}_{1}-\left( 1-y\right) 
\overrightarrow{q}_{2}\right)  \notag \\
&&\times \left( -\overrightarrow{v}_{12}\cdot \widehat{q}_{12}-sgn\left( 
\overrightarrow{v}_{12}\cdot \widehat{q}_{12}\right) \sqrt{\left( 
\overrightarrow{v}_{12}\cdot \widehat{q}_{12}\right) ^{2}-\frac{4}{m}\Delta
_{a}\left( \widehat{q}_{12}\cdot \overrightarrow{v}_{12}\right) }\right) . 
\notag
\end{eqnarray}%
Similarly, the heat flux has a kinetic contribution%
\begin{equation}
\overrightarrow{q}^{K}\left( \overrightarrow{r},t\right) =\frac{1}{2}m\int d%
\overrightarrow{v}_{1}\;f\left( \overrightarrow{r},\overrightarrow{v}%
_{1},t\right) \overrightarrow{V}_{1}V_{1}^{2}  \label{flux3}
\end{equation}%
and a collisional contribution%
\begin{eqnarray}
\overrightarrow{q}^{C}\left( \overrightarrow{r},t\right) &=&-\frac{m}{4V}%
\sigma \sum_{a}\int dx_{1}dx_{2}\;\widehat{q}_{12}\left( \widehat{q}%
_{12}\cdot \overrightarrow{v}_{12}\right) \delta \left( q_{12}-\sigma
\right) \Theta \left( -\widehat{q}_{12}\cdot \overrightarrow{v}_{12}\right)
\label{flux4} \\
&&\times \chi \left( \overrightarrow{q}_{1},\overrightarrow{q}_{2};\left[ n%
\right] \right) f\left( x_{1};t\right) f\left( x_{2};t\right) K_{a}\left( 
\widehat{q}_{12}\cdot \overrightarrow{v}_{12}\right) \int_{0}^{1}dx\;\delta
\left( \overrightarrow{r}-x\overrightarrow{q}_{1}-\left( 1-x\right) 
\overrightarrow{q}_{2}\right)  \notag \\
&&\times \frac{1}{2}\left( \overrightarrow{V}_{1}+\overrightarrow{V}%
_{2}\right) \cdot \widehat{q}_{12}\left( -\overrightarrow{v}_{12}\cdot 
\widehat{q}_{12}-sgn\left( \overrightarrow{v}_{12}\cdot \widehat{q}%
_{12}\right) \sqrt{\left( \overrightarrow{v}_{12}\cdot \widehat{q}%
_{12}\right) ^{2}-\frac{4}{m}\Delta _{a}\left( \widehat{q}_{12}\cdot 
\overrightarrow{v}_{12}\right) }\right) .  \notag
\end{eqnarray}%
Finally, because of the possibility of energy loss, the equation for the
temperature includes a source term given by%
\begin{eqnarray}
\xi \left( \overrightarrow{r},t\right) &=&\frac{1}{2V}\sum_{a}\int
dx_{1}dx_{2}\;\left( \widehat{q}_{12}\cdot \overrightarrow{v}_{12}\right)
\delta \left( q_{12}-\sigma \right) \Theta \left( -\widehat{q}_{12}\cdot 
\overrightarrow{v}_{12}\right)  \label{flux5} \\
&&\times K_{a}\left( \widehat{q}_{12}\cdot \overrightarrow{v}_{12}\right)
\Delta _{a}\left( \widehat{q}_{12}\cdot \overrightarrow{v}_{12}\right) \chi
\left( \overrightarrow{q}_{1},\overrightarrow{q}_{2};\left[ n\right] \right)
f\left( x_{1};t\right) f\left( x_{2};t\right) K_{a}\left( x_{12}\right)
\delta \left( \overrightarrow{r}-\overrightarrow{q}_{1}\right) .  \notag
\end{eqnarray}%
All of these expressions are exact, given the Enskog approximation, and show
that the hydrodynamics of the system is completely specified once the
one-body distribution is known.

\subsection{Chapman-Enskog expansion}

The Chapman-Enskog expansion is basically a gradient expansion of the
kinetic equation assuming a particular form for the solution. Specifically,
one attempts to construct a so-called normal solution in which all space and
time dependence occurs through the hydrodynamic fields%
\begin{equation}
f\left( \overrightarrow{r},\overrightarrow{v};t\right) =f\left( 
\overrightarrow{v}|\overrightarrow{r},\psi _{t}\right)
\end{equation}%
where the compact notation for the set of hydrodynamic fields $\psi
_{t}\left( \overrightarrow{r}\right) =\left\{ n\left( \overrightarrow{r}%
,t\right) ,T\left( \overrightarrow{r},t\right) ,\overrightarrow{u}\left( 
\overrightarrow{r},t\right) \right\} $ has been introduced and the notation
indicates that the distribution is a functional of the hydrodynamic fields
at time $t$. This means that time derivatives will be evaluated as 
\begin{equation}
\frac{\partial }{\partial t}f\left( \overrightarrow{r},\overrightarrow{v}%
;t\right) =\sum_{i}\int d\overrightarrow{r}^{\prime }\frac{\partial \psi
_{t,i}\left( \overrightarrow{r}^{\prime }\right) }{\partial t}\;\frac{\delta 
}{\delta \psi _{t,i}\left( \overrightarrow{r}^{\prime }\right) }f\left( 
\overrightarrow{r},\overrightarrow{v};t\right) .  \label{time-deriv}
\end{equation}%
To order the terms in the gradient expansion, we introduce a uniformity
parameter $\epsilon $ and replace $\overrightarrow{\nabla }$ with $\epsilon 
\overrightarrow{\nabla }$ and order terms in $\epsilon $. Since the space
and time derivatives are related by the balance equations, we also introduce
an expansion of the time derivative $\frac{\partial }{\partial t}=\partial
_{t}^{(0)}+\epsilon \partial _{t}^{(1)}+...$ as well as of the distribution
itself 
\begin{equation}
f\left( \overrightarrow{q}_{1},\overrightarrow{v}_{1},t\right) =f_{0}\left[ 
\overrightarrow{v}_{1}|\overrightarrow{q}_{1},\psi _{t}\right] +\epsilon
f_{1}\left[ \overrightarrow{v}_{1}|\overrightarrow{q}_{1},\psi _{t}\right]
+...
\end{equation}%
Finally, in the Enskog approximation the collision operator is nonlocal and
so must also be expanded (see Appendix \ref{AppExpandOperator}) as $J\left[
f,f\right] =J_{0}\left[ f,f\right] +\epsilon J_{1}\left[ f,f\right] +...$.
Substituting these expansions into the Enskog equation and equating terms
order by order in $\epsilon $gives a set of equations for the distribution
the first two of which are%
\begin{eqnarray}
\partial _{t}^{(0)}f_{0}\left( x_{1};t\right) &=&J_{0}\left[ f_{0},f_{0}%
\right]  \label{expansion} \\
\left( \partial _{t}^{(1)}+\overrightarrow{v}_{1}\cdot \frac{\partial }{%
\partial \overrightarrow{q}_{1}}+\right) f_{0}\left( x_{1};t\right)
+\partial _{t}^{(0)}f_{1}\left( x_{1};t\right) &=&J_{0}\left[ f_{0},f_{1}%
\right] +J_{0}\left[ f_{1},f_{0}\right] +J_{1}\left[ f_{0},f_{0}\right] . 
\notag
\end{eqnarray}%
Since all time and space dependence of the distribution occurs via the
hydrodynamic fields, the balance equations must also be expanded giving at
zeroth order%
\begin{eqnarray}
\partial _{t}^{(0)}n &=&0  \label{f0} \\
\partial _{t}^{(0)}n\overrightarrow{u} &=&0  \notag \\
\partial _{t}^{(0)}T &=&\frac{2}{Dnk_{B}}\xi ^{(0)}  \notag
\end{eqnarray}%
and at first order%
\begin{eqnarray}
\partial _{t}^{(1)}\rho +\overrightarrow{\nabla }\cdot \left( 
\overrightarrow{u}\rho \right) &=&0  \label{f1} \\
\partial _{t}^{(1)}\rho \overrightarrow{u}+\overrightarrow{\nabla }\cdot
\left( \rho \overrightarrow{u}\overrightarrow{u}\right) +\overrightarrow{%
\nabla }\cdot \overleftrightarrow{P}^{(0)} &=&0  \notag \\
\left( \partial _{t}^{(1)}+\overrightarrow{u}\cdot \overrightarrow{\nabla }%
\right) T+\frac{2}{Dnk_{B}}\left[ \overleftrightarrow{P}^{(0)}:%
\overrightarrow{\nabla }\overrightarrow{u}+\overrightarrow{\nabla }\cdot 
\overrightarrow{q}^{(0)}\right] &=&\frac{2}{Dnk_{B}}\xi ^{(1)}  \notag
\end{eqnarray}%
where, as noted, the fluxes and sources must also be expanded accordingly
(see Appendix \ref{AppExpandFluxes}). The logic of the normal solution is
that these balance equations define the meaning of the time derivatives so
that the time derivatives in eq.(\ref{expansion}) are evaluated using eq.(%
\ref{f1}) and eq.(\ref{time-deriv}). Together with the expressions for the
fluxes, eqs.(\ref{flux1})-(\ref{flux5}) suitably expanded, this gives a
closed set of integro-differential equations for the distribution function.

\section{Chapman-Enskog at zeroth order: the Homogeneous Cooling State}

\subsection{Expansion of the zeroth-order distribution}

At zeroth order in the gradient expansion, the distribution $f_{0}\left(
x_{1};t\right) $ must be a local function of the hydrodynamic fields so eqs.(%
\ref{expansion}) and (\ref{f0}) give%
\begin{equation}
\left( \frac{2}{Dnk_{B}}\xi ^{(0)}\right) \frac{\partial }{\partial T}%
f_{0}\left( x_{1};t\right) =J_{0}\left[ f_{0},f_{0}\right]  \label{ce0}
\end{equation}%
with 
\begin{eqnarray}
\xi ^{(0)}\left( \overrightarrow{r},t\right) &=&\frac{1}{2}\sum_{a}\int d%
\overrightarrow{v}_{1}d\overrightarrow{v}_{2}d\widehat{q}\;\left( \widehat{q}%
\cdot \overrightarrow{v}_{12}\right) \Theta \left( -\widehat{q}\cdot 
\overrightarrow{v}_{12}\right) K_{a}\left( \sigma \widehat{q}\cdot 
\overrightarrow{v}_{12}\right) \Delta _{a}\left( \sigma \overrightarrow{q},%
\overrightarrow{v}_{12}\right)  \label{s0} \\
&&\times \chi _{0}\left( \sigma ;n\left( \overrightarrow{r}\right) \right)
f_{0}\left( \overrightarrow{r},\overrightarrow{v}_{1};t\right) f_{0}\left( 
\overrightarrow{r},\overrightarrow{v}_{2};t\right) .  \notag
\end{eqnarray}%
where $\chi _{0}\left( \sigma ;n\right) $ is the pair distribution function
for a uniform equilibrium fluid of density $n$. Notice that this is not only
the zeroth order component of the Chapman-Enskog expansion, but that it is
also an exact (within the Enskog approximation) equation for the
distribution of a spatially homogeneous system. If there is no energy loss,
the solution will simply be the Maxwell distribution. When energy is lost in
collisions, and in the absence of external forcings, the system cools and
this is commonly known as the Homogeneous Cooling Solution (HCS).

To solve for the HCS, we expand the velocity dependence about an equilibrium
distribution by writing it as%
\begin{equation}
f_{0}(x_{1})=f_{M}\left( v_{1};\psi _{t}\right) \sum_{i}c_{i}\left( \psi
_{t}\right) S_{i}\left( \frac{m}{2k_{B}T}v_{1}^{2}\right)
\label{hsc-distribution}
\end{equation}%
where the Maxwellian distribution $f_{M}\left( v_{1};\psi _{t}\right) =n\pi
^{-D/2}\left( \frac{2k_{B}T}{m}\right) ^{-D/2}\exp \left( -\frac{m}{2k_{B}T}%
v_{1}^{2}\right) $ and it is important to note that this depends on the
exact local fields $\psi _{t}(\overrightarrow{r})$. The functions $\left\{
S_{i}\left( x\right) \right\} _{i=0}^{\infty }$ comprise a complete set of
polynomials which are orthogonal under a Gaussian measure so that%
\begin{equation}
\int d\overrightarrow{v}\;f_{M}\left( v_{1};\psi _{t}\right) S_{i}\left( 
\frac{m}{2k_{B}T}v_{1}^{2}\right) S_{i}\left( \frac{m}{2k_{B}T}%
v_{1}^{2}\right) =A_{i}\delta _{ij}  \label{6}
\end{equation}%
where $A_{i}$ is a normalization constant. In fact, these can be written in
terms of the Sonine, or associated Laguerre, polynomials%
\begin{equation}
L_{k}^{\alpha }\left( x\right) =\sum_{m=0}^{k}\frac{\Gamma \left( \alpha
+k+1\right) \left( -x\right) ^{m}}{\Gamma \left( \alpha +m+1\right) \left(
k-m\right) !m!}  \label{7}
\end{equation}%
and satisfy%
\begin{equation}
\int_{0}^{\infty }dx\;x^{\alpha }Lk\left( x\right) L_{m}^{\alpha }\left(
x\right) \exp \left( -x\right) =\frac{\Gamma \left( \alpha +k+1\right) }{%
\Gamma \left( k+1\right) }\delta _{mk}  \label{8}
\end{equation}%
so that in $D$ dimensions 
\begin{equation}
S_{k}\left( x\right) =L_{k}^{\frac{D-2}{2}}\left( x\right)  \label{9}
\end{equation}%
and%
\begin{equation}
A_{k}=\frac{\Gamma \left( \frac{1}{2}D+k\right) }{\Gamma \left( \frac{1}{2}%
D\right) \Gamma \left( k+1\right) }.
\end{equation}%
Substituting eq.(\ref{hsc-distribution}) into the differential equation, eq.(%
\ref{ce0}), multiplying through by $L_{k}^{\frac{D-2}{2}}\left( \frac{m}{%
2k_{B}T}v_{1}^{2}\right) $ and integrating gives%
\begin{equation}
\left( \frac{2}{Dnk_{B}T}\xi ^{(0)}\left( \psi _{t}\right) \right) \left[ T%
\frac{\partial }{\partial T}c_{k}\left( \psi _{t}\right) +k\left(
c_{k}\left( \psi _{t}\right) -c_{k-1}\left( \psi _{t}\right) \right) \right]
=\sum_{rs}I_{k,rs}\left( \psi _{t}\right) c_{r}\left( \psi _{t}\right)
c_{s}\left( \psi _{t}\right)  \label{a0}
\end{equation}%
with%
\begin{eqnarray}
I_{k,rs}\left( \psi _{t}\right) &=&-n^{-1}A_{k}^{-1}\int d\overrightarrow{v}%
_{1}d2\;L_{k}^{\frac{D-2}{2}}\left( \frac{m}{2k_{B}T}v_{1}^{2}\right) 
\overline{T}_{-}\left( 12\right) \left( \frac{2k_{B}T}{m}\right)
^{-D}f_{M}\left( v_{1};\psi _{t}\right)  \label{13} \\
&&\times f_{M}\left( v_{2};\psi _{t}\right) L_{r}^{\frac{D-2}{2}}\left( 
\frac{m}{2k_{B}T}v_{1}^{2}\right) L_{s}^{\frac{D-2}{2}}\left( \frac{m}{%
2k_{B}T}v_{1}^{2}\right) .  \notag
\end{eqnarray}

Notice that since $\xi ^{(0)}\sim n^{2}$ and $I_{k,rs}\sim n$, the
coefficients $c_{k}$ can only depend on temperature so $c_{k}\left( \psi
_{t}\right) =c_{k}\left( T(t)\right) $. Furthermore, it must be the case
that $c_{0}=1$ and $c_{1}=0$ in order to satisfy the definitions of the
hydrodynamic fields. It is easy to show that $I_{rs}^{0}=0$ so that the $k=0$
equation is trivial. Suppressing the dependence on $\psi _{t}$, the $k=1$
equation gives%
\begin{equation*}
-\frac{2}{Dnk_{B}}\xi ^{(0)}=\sum_{rs}I_{1,rs}c_{r}c_{s}
\end{equation*}%
and it may be confirmed that this is consistent with eq.(\ref{s0}). The
first nontrivial approximation is to take $c_{k}=0$ for $k>2$ and to use the 
$k=2$ equation to get 
\begin{eqnarray}
-\frac{2}{Dnk_{B}}\xi ^{(0)} &=&I_{1,00}+\left( I_{1,20}+I_{1,02}\right)
c_{2}  \label{a2} \\
\left( \frac{2}{Dnk_{B}T}\xi ^{(0)}\right) \left[ T\frac{\partial }{\partial
T}c_{2}+2c_{2}\right]  &=&I_{2,00}+\left( I_{2,20}+I_{2,02}\right) c_{2}
\end{eqnarray}%
where the quadratic terms on the right are typically neglected as they are
of similar structure to the neglected quartic terms. For a granular fluid,
there is no quantity with the units of energy except for the temperature, so
the coefficients of the expansion, which are dimensionless, are
temperature-independent. For systems with additional energy scales, the
coefficients must be determined by solving the resultant ordinary
differential equations with appropriate boundary conditions. For example, if
the energy loss were bounded, then at high temperatures it should be
insignificant and one would expect $\lim_{T\rightarrow \infty }c_{k}\left(
T\right) =\delta _{k0}$ to be the boundary condition.

\subsection{Generating function formalism}

The calculation of the integrals which define the coefficients on the left
in eq.(\ref{a0}) can be formulated in terms of a generating function.
Specifically, it is shown in Appendix \ref{AppGeneratingFunction} that 
\begin{eqnarray}
I_{k,rs}\left( \psi _{t}\right) &=&-n^{\ast }\frac{\Gamma \left( \frac{1}{2}%
D\right) \Gamma \left( k+1\right) }{\Gamma \left( \frac{1}{2}D+k\right) }%
\left( \frac{2k_{B}T}{m\sigma ^{2}}\right) ^{1/2}  \label{I} \\
&&\times \frac{1}{r!s!k!}\lim_{z_{1}\rightarrow 0}\lim_{z_{2}\rightarrow
0}\lim_{x\rightarrow 0}\frac{\partial ^{r}}{\partial z_{1}^{r}}\frac{%
\partial ^{s}}{\partial z_{2}^{s}}\frac{\partial ^{k}}{\partial x^{k}}\left[
\sum_{a}G_{a}\left( \psi _{t}|\Delta _{a}\right) -G_{0}\right]  \notag
\end{eqnarray}%
where $n^{\ast }=n\sigma ^{D}.$The generating functions are 
\begin{eqnarray}
G_{a}\left( \psi _{t}|\Delta _{a}\right) &=&-\frac{1}{2}\pi
^{-1/2}S_{D}\left( 1-z_{1}x\right) ^{-\frac{1}{2}D}\allowbreak \left( \frac{%
1-z_{1}x}{2-x-z_{2}-z_{1}+xz_{1}z_{2}}\right) ^{\frac{1}{2}} \\
&&\times \int_{0}^{\infty }du\;K_{a}^{\ast }\left( \sqrt{u}\right) \exp
\left( \frac{\left( 2-z_{2}-z_{1}\right) x}{2-x-z_{2}-z_{1}+xz_{1}z_{2}}%
\frac{1}{2}\Delta _{a}^{\ast }\left( \sqrt{u}\right) \right)  \notag \\
&&\times \exp \left( -\frac{1-z_{2}x}{2-x-z_{2}-z_{1}+xz_{1}z_{2}}u\right) 
\notag \\
&&\times \exp \left( -\frac{1}{2}\frac{\left( z_{2}-z_{1}\right) x}{%
2-x-z_{2}-z_{1}+xz_{1}z_{2}}\left( u-\sqrt{u}\sqrt{u-2\Delta _{a}^{\ast
}\left( \sqrt{u}\right) }\right) \right)  \notag
\end{eqnarray}%
and%
\begin{equation}
G_{0}=-\frac{1}{2}\pi ^{-1/2}S_{D}\left( 1-z_{1}x\right) ^{-\frac{D+1}{2}%
}\left( 2-x-z_{2}-z_{1}+xz_{1}z_{2}\right) ^{\frac{1}{2}}.
\end{equation}%
with $S_{D}$ the area of the $D$ dimensional unit hard sphere, 
\begin{equation}
S_{D}=\frac{2\pi ^{D/2}}{\Gamma \left( D/2\right) }
\end{equation}%
which is, e.g., $4\pi $ in three dimensions and $2\pi $ for $D=2$. The
scaled probabilities and energy loss functions are 
\begin{eqnarray}
K^{\ast }\left( x\right) &=&K\left( x\sqrt{\frac{2}{m\beta }}\right) \\
\Delta ^{\ast }\left( x\right) &=&\beta \Delta \left( x\sqrt{\frac{2}{m\beta 
}}\right) .  \notag
\end{eqnarray}%
In general, $K^{\ast }\left( x\right) $ and $\Delta ^{\ast }\left( x\right) $
are functions of temperature and, hence, time but in order to keep the
resulting expressions below from becoming too cumbersome, these arguments
will be suppressed.The utility of this generating function, which is
admittedly complex, is that the limits and derivatives needed to evaluate
eq.(\ref{I}) are easily programmed using symbolic manipulation packages.

To complete the description of the uniform fluid, I give the quantities
necessary to calculate the lowest order correction. These are written
conveniently as 
\begin{eqnarray}
I_{k,rs} &=&n^{\ast }\chi \frac{S_{D}}{2D\left( D+2\right) \sqrt{\pi }}%
\left( \frac{k_{B}T}{m\sigma ^{2}}\right) ^{1/2}I_{k,rs}^{\ast } \\
I_{k,rs}^{\ast } &=&I_{k,rs}^{\ast E}+\sum_{a}\int_{0}^{\infty }K_{a}\left( 
\sqrt{u}\right) I_{k,rs}^{\ast I}e^{-\frac{1}{2}u}du  \notag
\end{eqnarray}%
where $I_{02,2}^{\ast E}+I_{20,2}^{\ast E}=-8\left( D-1\right) $ and all
other elastic contributions are zero, and the inelastic kernals are 
\begin{eqnarray}
I_{1,00}^{\ast I} &=&\left( D+2\right) \Delta ^{\ast }\left( \sqrt{u}\right)
\\
I_{2,00}^{\ast I} &=&\left( \Delta ^{\ast }\left( \sqrt{u}\right)
+3-u\right) \Delta ^{\ast }\left( \sqrt{u}\right)  \notag \\
I_{1,02}^{\ast I}+I_{1,20}^{\ast I} &=&\frac{D+2}{16}\left(
u^{2}-6u+3\right) \Delta ^{\ast }\left( \sqrt{u}\right)  \notag \\
I_{02,2}^{\ast I}+I_{20,2}^{\ast I} &=&\frac{1}{16}\left( 
\begin{array}{c}
\Delta ^{\ast }\left( \sqrt{u}\right) \left( \Delta ^{\ast }\left( \sqrt{u}%
\right) \left( u^{2}-6u+3\right) -u^{3}+9u^{2}-\left( 8D+49\right)
u+8D+37\right) \allowbreak \\ 
+16\left( D-1\right) \left( u-\sqrt{u}\sqrt{\left( u-2\Delta ^{\ast }\left( 
\sqrt{u}\right) \right) }\right)%
\end{array}%
\right)  \notag
\end{eqnarray}

The pressure can similarly be expressed in the generating function
formalism, but here we just give for later use the expression for the
pressure (see Appendix \ref{AppExpandFluxes}) including the lowest order
corrections to the Gaussian distribution 
\begin{eqnarray}
\frac{p^{\left( 0\right) }}{nk_{B}T} &=&1+n^{\ast }\chi \frac{S_{D}}{2D}
\label{pressure} \\
&&+nk_{B}Tn^{\ast }\chi \frac{S_{D}}{2D}\frac{1}{\sqrt{2\pi }}%
\sum_{a}\int_{0}^{\infty }vK_{a}^{\ast }\left( -v\right) g\left( v,\Delta
_{a}^{\ast }\left( -v\right) \right) \left( 1+\frac{D}{16}c_{2}\left(
v^{4}-6v^{2}+3\right) \right) \exp \left( -\frac{1}{2}v^{2}\right) dv. 
\notag
\end{eqnarray}%
where the function $g\left( v,\Delta \right) $ is defined as%
\begin{equation*}
g\left( v,\Delta \right) =sgn\left( v\right) \sqrt{v^{2}-2\Delta }-v.
\end{equation*}

\subsection{The simple granular fluid in D dimensions}

For a simple granular fluid with constant coefficient of restitution one has 
$\Delta ^{\ast }\left( v\right) =\beta \Delta \left( v\sqrt{\frac{2k_{B}T}{m}%
}\right) =\left( 1-\alpha ^{2}\right) \frac{1}{2}v^{2}$ and $K^{\ast }\left(
x\right) =1$ giving

\begin{eqnarray}
G\left( \psi _{t}|\Delta \right) &\rightarrow &-\pi ^{-1/2}S_{D}\left(
1-z_{1}x\right) ^{-\frac{1}{2}D}\allowbreak \left( 1-z_{1}x\right) ^{\frac{1%
}{2}}\left( 2-x-z_{2}-z_{1}+xz_{1}z_{2}\right) ^{1/2} \\
&&\times \left[ -\frac{1}{2}\left( 2-z_{2}-z_{1}\right) x\left( 1-\alpha
^{2}\right) +2-2z_{2}x+\left( z_{2}-z_{1}\right) x\left( 1-\alpha \right) %
\right] ^{-1}.  \notag
\end{eqnarray}%
and 
\begin{eqnarray}
I_{1,00}^{\ast } &=&2\left( D+2\right) \left( 1-\alpha ^{2}\right) \\
I_{2,00}^{\ast } &=&\allowbreak 2\left( 1-2\alpha ^{2}\right) \left(
1-\alpha ^{2}\right)  \notag \\
I_{1,02}^{\ast }+I_{1,20}^{\ast } &=&\frac{3\left( D+2\right) }{8}\left(
1-\alpha ^{2}\right)  \notag \\
I_{2,02}^{\ast }+I_{2,20}^{\ast } &=&-8\left( D-1\right) +\frac{1}{8}\left(
\alpha -1\right) \left( 30\alpha ^{3}+30\alpha ^{2}+24D\alpha +105\alpha
+137-8D\right)  \notag \\
&=&\allowbreak \allowbreak \frac{1}{8}\left( \alpha +1\right) \left(
30\alpha ^{3}-30\alpha ^{2}+24D\alpha +105\alpha -56D-73\right) \allowbreak 
\notag
\end{eqnarray}%
so that the lowest order correction to the Gaussian is 
\begin{eqnarray}
c_{2} &=&\frac{I_{2,00}}{-2I_{1,00}-\left( I_{2,20}+I_{2,02}\right) } \\
&=&\frac{16\left( 1-2\alpha ^{2}\right) \left( 1-\alpha \right) }{%
24D+9-\left( 41-8D\right) \alpha +30\alpha ^{2}\left( 1-\alpha \right) } 
\notag
\end{eqnarray}%
The cooling rate is%
\begin{equation}
\xi ^{(0)}=-\left( 1-\alpha ^{2}\right) n^{\ast }\chi \frac{S_{D}}{2\sqrt{%
\pi }}\left( \frac{k_{B}T}{m\sigma ^{2}}\right) ^{1/2}nk_{B}T\left[ 1+\frac{3%
}{16}c_{2}\right]  \label{s1}
\end{equation}%
and for a simple granular fluid, $g\left( v,\Delta ^{\ast }\left( -v\right)
\right) =v\left( \alpha -1\right) $ gives 
\begin{equation}
p^{\left( 0\right) }=nk_{B}T\left\{ 1+n^{\ast }\chi \frac{S_{D}}{4D}\left(
1+\alpha \right) \right\}
\end{equation}%
so it is seen that the second order terms do not contribute.

This completes the discussion of the zeroth order Chapman-Enskog solution
which is also the HCS. The simplest approximation is to take the
distribution to be Maxwellian with the temperature obeying eq.(\ref{f0}).
The first nontrivial approximation is to keep $c_{2}$ which is given by eq.(%
\ref{a2}) with coefficients that depend on the energy loss model.

\section{Chapman-Enskog at first order: the Navier-Stokes equations}

The first order equation can be written as%
\begin{equation}
\partial _{t}^{(0)}f_{1}-\mathcal{L}_{0}\left[ f_{1}\right] =J_{1}\left[
f_{0},f_{0}\right] -\left( \partial _{t}^{(1)}+\overrightarrow{v}\cdot 
\overrightarrow{\nabla }\right) f_{0}
\end{equation}%
where $\mathcal{L}_{0}\left[ f_{1}\right] =\left( J_{0}\left[ f_{1},f_{0}%
\right] +J_{0}\left[ f_{0},f_{1}\right] \right) $ is the linearized
Boltzmann operator. The first order balance equations are used to eliminate
the time derivative on the right. It is convenient to divide the first order
heat source into two parts%
\begin{equation}
\xi _{1}=\xi _{0}\left[ f_{1}\right] +\xi _{1}\left[ f_{0}\right]
\end{equation}%
where the first term on the right is, as indicated, a linear operator acting
on the first order distribution and the second is of first order in the
gradients and depends solely on the zeroth order distribution. Furthermore,
since it is a scalar, $\xi _{1}\left[ f_{0}\right] $ must be proportional to
the only scalar gradient, namely $\overrightarrow{\nabla }\cdot 
\overrightarrow{u}$ so that we will write $\xi _{1}\left[ f_{0}\right] =$ $%
\xi _{1}^{\nabla u}\left[ f_{0}\right] \overrightarrow{\nabla }\cdot 
\overrightarrow{u}$ (see appendix \ref{AppExpandFluxes} for more details).
Then, the first order equation becomes%
\begin{equation}
\partial _{t}^{(0)}f_{1}+\frac{2}{Dnk_{B}T}\xi _{0}\left[ f_{1}\right]
\left( T\frac{\partial }{\partial T}f_{0}\right) -\mathcal{L}_{0}\left[ f_{1}%
\right] =J_{1}\left[ f_{0},f_{0}\right] -\frac{2}{Dnk_{B}T}\xi _{1}^{\nabla
u}\left[ f_{0}\right] \left( \overrightarrow{\nabla }\cdot \overrightarrow{u}%
\right) \left( T\frac{\partial }{\partial T}f_{0}\right) -\sum_{\alpha
}\sum_{i}B_{i}^{\alpha }\left( \overrightarrow{V};\left[ f_{0}\right]
\right) \partial _{i}\psi _{t,\alpha }  \label{ce1}
\end{equation}%
with 
\begin{eqnarray}
B_{i}^{n} &=&\left( n^{-1}f_{0}+\frac{1}{nk_{B}T}\frac{\partial p^{(0)}}{%
\partial n}\left[ \frac{\partial }{\partial z_{1}}f_{0}\right] _{T}\right)
V_{1i} \\
B_{i}^{T} &=&\left( \frac{1}{nk_{B}T}\frac{\partial p^{(0)}}{\partial T}%
\left[ \frac{\partial }{\partial z_{1}}f_{0}\right] _{T}+\frac{\partial }{%
\partial T}f_{0}\right) V_{1i}  \notag \\
B_{i}^{u_{j}} &=&\frac{2}{Dnk_{B}}\left( -p^{(0)}\frac{\partial }{\partial T}%
f_{0}-\frac{mnV_{1}^{2}}{2T}\left[ \frac{\partial }{\partial z_{1}}f_{0}%
\right] _{T}-\frac{Dnk_{B}}{2}f_{0}\right) \delta _{ij}  \notag \\
&&+\frac{m}{2k_{B}T}\left( V_{1i}V_{1j}-\frac{1}{D}V_{1}^{2}\delta
_{ij}\right) \left[ -\frac{\partial }{\partial z_{1}}f_{0}\right] _{T} 
\notag
\end{eqnarray}%
where the variable $z=\frac{m}{2k_{B}T}V^{2}$.

It is shown in Appendix \ref{AppExpandOperator} that $J_{1}\left[ f_{0},f_{0}%
\right] $ can be written as%
\begin{equation}
J_{1}\left[ f_{0},f_{0}\right] =\sum_{\gamma ,i}\left( \partial _{i}\psi
_{t,\gamma }\left( \overrightarrow{r}\right) \right) \cdot \left( J_{i}^{(0)}%
\left[ f_{0},\frac{\partial }{\partial \psi _{\gamma }}f_{0}\right] +\frac{1%
}{2}\delta _{\gamma n}\frac{\partial \ln \chi }{\partial n}J_{i}^{(0)}\left[
f_{0},f_{0}\right] \right)
\end{equation}%
where the detailed form of the operator $J_{i}^{(0)}$ is given in the
Appendix. The right hand side of eq.(\ref{ce1}) is therefore expressed
entirely in terms of the gradients of the hydrodynamic fields so that the
first order correction to the distribution must also be proportional to the
gradients. Since the only vector available is $\overrightarrow{V}$ and the
only tensors are the unit tensor and the symmetric traceless tensor $%
V_{i}V_{j}-\frac{1}{D}\delta _{ij}V^{2}$, the first order distribution must
take the form%
\begin{equation}
f_{1}=f_{0}\left( x_{1}\right) \left[ 
\begin{array}{c}
A^{\left( n\right) }\left( \overrightarrow{V}_{1}\right) V_{1i}\partial
_{i}n+A^{\left( T\right) }\left( \overrightarrow{V}_{1}\right)
V_{1i}\partial _{i}T+A^{\left( \nabla u\right) }\left( \overrightarrow{V}%
_{1}\right) \overrightarrow{\nabla }\cdot \overrightarrow{u} \\ 
+\sqrt{\frac{D}{D-1}}A^{\left( \partial u\right) }\left( \overrightarrow{V}%
_{1}\right) \left( V_{1i}V_{1j}-\frac{1}{D}\delta _{ij}V^{2}\right) \left(
\partial _{i}u_{j}+\partial _{j}u_{i}-\frac{2}{D}\delta _{ij}\overrightarrow{%
\nabla }\cdot \overrightarrow{u}\right)%
\end{array}%
\right] .  \label{ff1}
\end{equation}%
Then, both sides of the kinetic equation are expressed in terms of the
gradients of the hydrodynamic fields and since those gradients can vary
independently, their coefficients must vanish individually giving%
\begin{align}
& \phi _{I_{\alpha }}^{\alpha }\left( \overrightarrow{V}_{1}\right) \left[ 
\frac{2}{Dnk_{B}T}\xi _{0}\left[ f_{0}\right] \frac{\partial }{\partial T}%
f_{0}A^{\left( \alpha \right) }+\sum_{\gamma }K_{\gamma }^{\alpha }\left[
A^{\left( \gamma \right) }\right] \right] -\mathcal{L}_{0}\left[
f_{0}A^{\left( \alpha \right) }\phi _{I_{\alpha }}^{\alpha }\right]
\label{XX} \\
& =\Omega _{I_{\alpha }}^{\alpha }\left[ f_{0},f_{0}\right] -C^{\alpha
}\left( \overrightarrow{V}|f_{0}\right) \phi _{I_{\alpha }}^{\alpha }\left( 
\overrightarrow{V}_{1}\right) ,  \notag
\end{align}%
where Greek indices range over the four values $n,T,\nabla u$ and $\partial
u $. In this equation, the capitalized index, $I_{\alpha }$, is a
super-index corresponding to a set of Cartesian indices as illustrated by
the definition 
\begin{equation}
\phi _{I_{\alpha }}^{\alpha }\left( \overrightarrow{V}\right) =\left(
V_{i},V_{i},1,\sqrt{\frac{D}{D-1}}\left( V_{i}V_{j}-\frac{1}{D}V^{2}\delta
_{ij}\right) \right) .
\end{equation}%
The linear functional $K_{\gamma }^{\alpha }$ encapsulates contributions
coming from the action of the functional derivative on the non-local term $%
\overrightarrow{\nabla }T$ as well as terms coming from $\xi ^{(0)}\left[
f_{1}\right] $ and is given by 
\begin{eqnarray}
K_{\gamma }^{\alpha }\left[ A^{\left( \gamma \right) }\right] &=&A^{(T)}f_{0}%
\left[ \delta _{\alpha n}\frac{\partial }{\partial n}+\delta _{\alpha T}%
\frac{\partial }{\partial T}\right] \left( \frac{2}{Dnk_{B}T}\xi _{0}\left[
f_{0}\right] \right) \\
&&-\delta _{\alpha \nabla u}\frac{1}{Dnk_{B}T}\xi _{0}\left[ f_{0}\right] %
\left[ f_{0}A^{(\nabla u)}\right] \frac{\partial }{\partial T}f_{0}  \notag
\end{eqnarray}%
The source terms on the right are, after some manipulation, given by%
\begin{eqnarray}
C^{n}\left( \overrightarrow{V}|f_{0}\right) &=&n^{-1}f_{0}+\frac{1}{nk_{B}T}%
\frac{\partial p^{(0)}}{\partial n}\left[ \frac{\partial }{\partial z}f_{0}%
\right] _{T} \\
C^{T}\left( \overrightarrow{V}|f_{0}\right) &=&\frac{1}{nk_{B}T}\frac{%
\partial p^{(0)}}{\partial T}\left[ \frac{\partial }{\partial z}f_{0}\right]
_{T}+\frac{\partial }{\partial T}f_{0}  \notag \\
C^{\nabla u}\left( \overrightarrow{V}|f_{0}\right) &=&-\frac{2}{Dnk_{B}T}%
\left( \xi _{1}^{\nabla u}\left[ f_{0}\right] -p^{(0)C}\right) \left( \frac{D%
}{2}f_{0}+\frac{mV^{2}}{2T}\left[ \frac{\partial }{\partial z}f_{0}\right]
_{T}\right)  \notag \\
C^{\partial u}\left( \overrightarrow{V}|f_{0}\right) &=&-\frac{m}{2k_{B}T}%
\left[ \frac{\partial }{\partial z}f_{0}\right] _{T}  \notag
\end{eqnarray}%
and%
\begin{eqnarray}
\Omega _{i}^{n}\left[ f_{0},f_{0}\right] &=&\frac{1}{2}\frac{\partial \ln
n^{2}\chi }{\partial n}J_{i}^{(0)}\left[ f_{0},f_{0}\right]  \label{Omega} \\
\Omega _{i}^{T}\left[ f_{0},f_{0}\right] &=&J_{i}^{(0)}\left[ f_{0},\frac{%
\partial }{\partial T}f_{0}\right]  \notag \\
\Omega ^{\nabla u}\left[ f_{0},f_{0}\right] &=&\frac{1}{D}\sum_{i}J_{i}^{(0)}%
\left[ f_{0},\frac{\partial }{\partial u_{i}}f_{0}\right]  \notag \\
\Omega _{ij}^{\partial u}\left[ f_{0},f_{0}\right] &=&\frac{1}{4}\left( 
\begin{array}{c}
J_{j}^{(0)}\left[ f_{0},\frac{\partial }{\partial u_{i}}f_{0}\right]
+J_{i}^{(0)}\left[ f_{0},\frac{\partial }{\partial u_{j}}f_{0}\right] \\ 
-\frac{2}{D}\delta _{ij}\sum_{l}J_{l}^{(0)}\left[ f_{0},\frac{\partial }{%
\partial u_{l}}f_{0}\right]%
\end{array}%
\right) .  \notag
\end{eqnarray}

We conclude the discussion of the first order approximation with some
general remarks concerning the solution of eqs.(\ref{XX})-(\ref{Omega}).
First, the hydrodynamic fields are defined by eq.(\ref{fields}) which can be
written as 
\begin{equation}
\psi _{t,i}\left( \overrightarrow{r}\right) =\int \widehat{\psi }_{i}\left( 
\overrightarrow{V}\right) f\left( \overrightarrow{r},\overrightarrow{V}%
;t\right) d\overrightarrow{V}  \label{f3}
\end{equation}%
with the array of velocity moments $\widehat{\psi }\left( \overrightarrow{V}%
\right) =\left( 1,\frac{m}{2}V^{2},\overrightarrow{V}\right) $. However,
from the definition eq.(\ref{hsc-distribution}), it is clear that the zeroth
order distribution satisfies%
\begin{equation}
\psi _{t,i}\left( \overrightarrow{r}\right) =\int \widehat{\psi }_{i}\left( 
\overrightarrow{V}\right) f_{0}\left( \overrightarrow{r},\overrightarrow{V}%
;t\right) d\overrightarrow{V}
\end{equation}%
so that it must be the case that all higher order contributions to the
distribution give%
\begin{equation}
0=\int \widehat{\psi }_{i}\left( \overrightarrow{V}\right) f_{j}\left( 
\overrightarrow{r},\overrightarrow{V};t\right) d\overrightarrow{V}
\end{equation}%
for all $i$ and $j$. Since the gradients of the hydrodynamic fields are
arbitrary, this means that in the case of the first order distribution, the
coefficients of the gradients must be orthogonal to the first three velocity
moments under the measure $f_{0}\left( \overrightarrow{V}\right) $ or%
\begin{equation}
0=\int \widehat{\psi }_{i}\left( \overrightarrow{V}\right) A^{\left( \alpha
\right) }\left( \overrightarrow{V}\right) f_{0}\left( \overrightarrow{r},%
\overrightarrow{V};t\right) d\overrightarrow{V}.  \label{orthog}
\end{equation}

Second, it is clear that eq.(\ref{XX}) is a linear equation in the
coefficients $A^{\left( \alpha \right) }\left( \overrightarrow{V}\right) $
so that the conditions for the existence of a solution follows the usual
theory of linear operators. In particular, defining a Hilbert space with
measure $f_{0}\left( \overrightarrow{V}\right) $ the Fredholm alternative,
which states that for linear operator $L$ and source term $B$, the equation $%
LV=B$ has a solution if and only if $B$ is orthogonal to the null space of $%
L $. We expect that $\widehat{\psi }\left( \overrightarrow{V}\right) $ is in
the null space of the operator defined by the right hand side of eq.(\ref{XX}%
). In fact, it is clear that multiplying by $\widehat{\psi }_{i}\left( 
\overrightarrow{V}_{1}\right) \;$and integrating over velocities gives, on
the right hand side, 
\begin{equation}
-\delta _{\alpha \nabla u}\frac{1}{Dnk_{B}T}\xi _{0}\left[ f_{0}A^{(\nabla
u)}\right] \frac{\partial }{\partial T}\int d\overrightarrow{V}_{1}\;%
\widehat{\psi }_{i}\left( \overrightarrow{V}\right) \phi _{I_{\alpha
}}^{\alpha }\left( \overrightarrow{V}_{1}\right) f_{0}-\int d\overrightarrow{%
V}_{1}\;\widehat{\psi }_{i}\left( \overrightarrow{V}_{1}\right) \mathcal{L}%
_{0}\left[ f_{0}A^{\left( \alpha \right) }\phi _{I_{\alpha }}^{\alpha }%
\right]
\end{equation}%
Now, the $\mathcal{L}_{0}$ term vanishes for $\;\widehat{\psi }_{i}=1$ and $%
\overrightarrow{V}$ due to the conservation of particle number and total
momentum respectively. For the last choice, $\widehat{\psi }_{i}=\frac{m}{2}%
V^{2}$, it is only nonzero for $\alpha =\nabla u$ due to rotational symmetry
(for other choices of $\alpha $, $\phi _{I_{\alpha }}^{\alpha }$ is a vector
or traceless tensor). Thus the only non-vanishing element of this system of
equations is that for $\alpha =\nabla u$ and $\widehat{\psi }_{i}=\frac{m}{2}%
V^{2}$ which becomes 
\begin{equation}
-\frac{1}{2T}\xi _{0}\left[ f_{0}A^{(\nabla u)}\right] -\int d%
\overrightarrow{V}_{1}\;\frac{m}{2}V^{2}\mathcal{L}_{0}\left[ f_{0}A^{\left(
\nabla u\right) }\right]
\end{equation}%
and which is seen to vanish from the definition of $\xi _{0}\left[ g\right] $%
, eq.(\ref{psi0}) and $\mathcal{L}_{0}\left[ g\right] $. Thus, $\widehat{%
\psi }$ is indeed in the null space of the linear operator and a necessary
condition for the existence of a solution is that the right hand side is
orthogonal to $\widehat{\psi }$ as well. That this is indeed the case is
easily verified.

\subsection{Approximate solution of the integral equations}

The integral equations summarized by eq.s(\ref{XX})-(\ref{Omega}) will be
solved by expanding the unknown functions $A^{\left( \gamma \right) }$ in
associated Laguerre polynomials as%
\begin{equation}
A^{\left( \alpha \right) }\left( \overrightarrow{V}\right)
=\sum_{s}a_{s}^{\left( \alpha \right) }L_{s}^{\left( \frac{D-2}{2}+\lambda
_{\alpha }\right) }\left( \frac{m}{2k_{B}T}V^{2}\right)  \label{cex}
\end{equation}%
where the coefficients are in general functions of the hydrodynamic fields, $%
a_{s}^{\left( \alpha \right) }=a_{s}^{\left( \alpha \right) }\left( \psi
_{t}\right) $ although, for clarity, this dependence will be suppressed
below. It is interesting to note that for a simple granular gas, Garzo and
Dufty\cite{DuftyGranularTransport} wrote the first order distribution as in
eq.(\ref{ff1}) but with $f_{0}$ replaced by the Maxwellian $f_{M}\left(
v_{1};\psi _{t}\right) $. The use of $f_{0}$ here is motivated by the fact
that the source term in the first order equations, eq.(\ref{XX}), is
proportional to $f_{0}$ so that it seems appropriate to use this in the
definition of the first order correction but this is not necessary. In fact,
one clearly has that%
\begin{eqnarray}
f_{0}\left( \overrightarrow{V}\right) A^{\left( \alpha \right) }\left( 
\overrightarrow{V}\right) &=&f_{0}\left( \overrightarrow{V}\right)
\sum_{s}a_{s}^{\left( \alpha \right) }L_{s}^{\left( \frac{D-2}{2}+\lambda
_{\alpha }\right) }\left( \frac{m}{2k_{B}T}V^{2}\right) \\
&=&f_{M}\left( V;\psi _{t}\right) \sum_{s}\overline{a}_{s}^{\left( \alpha
\right) }L_{s}^{\left( \frac{D-2}{2}+\lambda _{\alpha }\right) }\left( \frac{%
m}{2k_{B}T}V^{2}\right)  \notag
\end{eqnarray}%
for new coefficients $\overline{a}_{s}^{\left( \alpha \right) }$ which are
linear combinations of the $a_{s}^{\left( \alpha \right) }$ and in
particular, if $a_{s_{0}}^{\left( \alpha \right) }$ is the first
non-vanishing coefficient in the sum, then $a_{s_{0}}^{\left( \alpha \right)
}=\overline{a}_{s_{0}}^{\left( \alpha \right) }$. In the development to
follow, the usual approximation will be made wherein only the lowest
non-vanishing coefficient is retained (the ''lowest Sonine approximation'')\
so that the two choices should be equivalent. However, in this
approximation, the exact relation $a_{s_{0}}^{\left( \alpha \right) }=%
\overline{a}_{s_{0}}^{\left( \alpha \right) }$ is only preserved if terms of
the form $c_{2}a_{s_{0}}^{\left( \alpha \right) }$ are systematically
neglected. This therefore motivates the simplifying approximation made here
whereby such terms are in fact neglected throughout. In the Conclusions, the
present approximation is evaluated for the special case of a simple granular
fluid in 3 dimensions by comparison to expressions given in ref. \cite%
{DuftyGranularTransport} where all dependence on $c_{2}$ is retained.

In order to solve the integral equations, the expansions eq.(\ref{cex}) are
substituted into eq.(\ref{XX})and the $\alpha $-th equation is multiplied by 
$\phi _{I_{\alpha }}^{\alpha }\left( \overrightarrow{V}_{1}\right)
L_{k}^{\left( \frac{D-2}{2}+\lambda _{\alpha }\right) }\left( \frac{m}{%
2k_{B}T}V_{1}^{2}\right) $ and tensorial indices contracted and $%
\overrightarrow{V}_{1}$ integrated. The left hand side of these equations is
found to be simplified by the choices $\lambda _{n}=\lambda _{T}=1$, $%
\lambda _{\nabla u}=0$ and $\lambda _{\partial u}=2$ which are made
henceforth. The result after some simplification can be written as%
\begin{equation}
\frac{2}{Dk_{B}}\xi _{0}\frac{\Gamma \left( \frac{D}{2}+\lambda _{\alpha
}+k\right) }{\Gamma \left( D/2\right) \Gamma \left( k+1\right) }\left( \frac{%
2k_{B}T}{m}\right) ^{\lambda _{\alpha }}\left[ \frac{\partial }{\partial T}%
a_{k}^{\left( \alpha \right) }+\frac{1}{T}\left( \lambda _{\alpha }+k\right)
a_{k}^{\left( \alpha \right) }-\frac{1}{T}ka_{k-1}^{\left( \alpha \right) }%
\right] +\sum_{l}I_{kl}^{\alpha }a_{l}^{\left( \alpha \right)
}-\sum_{l}K_{kl}^{\alpha \alpha ^{\prime }}a_{l}^{\left( \alpha ^{\prime
}\right) }=\Lambda _{k}^{\alpha }+\Omega _{k}^{\alpha }
\end{equation}%
where the contributions from the Boltzmann operator are%
\begin{equation}
I_{kl}^{\alpha }=-\sum_{I_{\alpha }}\int d\overrightarrow{V}_{1}\;\phi
_{I_{\alpha }}^{\alpha }\left( \overrightarrow{V}_{1}\right) L_{k}^{\left( 
\frac{D-2}{2}+\lambda _{\alpha }\right) }\left( \frac{m}{2k_{B}T}%
V_{1}^{2}\right) \mathcal{L}_{0}\left[ L_{l}^{\left( \frac{D-2}{2}+\lambda
_{\alpha }\right) }\left( \frac{m}{2k_{B}T}V^{2}\right) \phi _{I_{\alpha
}}^{\alpha }\left( \overrightarrow{V}\right) \right] ,
\end{equation}%
and the last coefficient on the left is%
\begin{equation}
K_{kl}^{\alpha \alpha ^{\prime }}=\int d\overrightarrow{V}_{1}\;\phi
_{I_{\alpha }}\left( \overrightarrow{V}_{1}\right) L_{k}^{\left( \frac{D-2}{2%
}+\lambda _{\alpha }\right) }\left( \frac{m}{2k_{B}T}V_{1}^{2}\right)
K_{\alpha ^{\prime }}^{\alpha }\left[ \phi _{I_{\alpha ^{\prime }}}\left( 
\overrightarrow{V}_{1}\right) L_{l}^{\left( \frac{D-2}{2}+\lambda _{\alpha
^{\prime }}\right) }\left( \frac{m}{2k_{B}T}V^{2}\right) \right]
\end{equation}%
The source terms on the right are%
\begin{equation}
\Lambda _{k}^{\alpha }=-\int d\overrightarrow{V}_{1}\;\phi _{I_{\alpha
}}\left( \overrightarrow{V}_{1}\right) L_{k}^{\left( \frac{D-2}{2}+\lambda
_{\alpha }\right) }\left( \frac{m}{2k_{B}T}V_{1}^{2}\right) C^{\alpha
}\left( \overrightarrow{V}|f_{0}\right) \phi _{I_{\alpha }}\left( 
\overrightarrow{V}_{1}\right)
\end{equation}%
and%
\begin{equation}
\Omega _{k}^{\alpha }=\int d\overrightarrow{V}_{1}\;\phi _{I_{\alpha
}}\left( \overrightarrow{V}_{1}\right) L_{k}^{\left( \frac{D-2}{2}+\lambda
_{\alpha }\right) }\left( \frac{m}{2k_{B}T}V_{1}^{2}\right) \Omega
_{i}^{\alpha }\left[ f_{0},f_{0}\right] .
\end{equation}

A straightforward evaluation using the orthogonality and standard recursion
relations of the associated Laguerre polynomials\cite{AbramStegun} gives%
\begin{eqnarray}
K_{kl}^{\alpha \alpha ^{\prime }} &=&\frac{\Gamma \left( \frac{1}{2}%
D+k+1\right) }{\Gamma \left( \frac{1}{2}D\right) \Gamma \left( k+1\right) }%
n\left( \frac{2k_{B}T}{m}\right) \delta _{\alpha ^{\prime }T}\delta
_{kl}\left( \delta _{\alpha n}\frac{\partial }{\partial n}+\delta _{\alpha T}%
\frac{\partial }{\partial T}\right) \frac{2}{Dnk_{B}}\xi _{0} \\
&&+\delta _{k1}\delta _{\alpha \nabla u}\delta _{\alpha ^{\prime }\nabla u}%
\frac{1}{2nk_{B}T}\xi ^{(0)}\left[ f_{0}L_{l}^{\left( \frac{D-2}{2}\right)
}\left( \frac{m}{2k_{B}T}V^{2}\right) \right] \left( \frac{1}{T}\right) n 
\notag
\end{eqnarray}%
and 
\begin{equation}
\Lambda _{k}^{\alpha }=\frac{\Gamma \left( \frac{D-2}{2}+k+1\right) }{\Gamma
\left( \frac{D}{2}\right) \Gamma \left( k+1\right) }\left( 
\begin{array}{c}
\frac{2k_{B}T}{m}\left( \frac{1}{2}D+k\right) \left[ \frac{1}{k_{B}T}\frac{%
\partial p^{(0)C}}{\partial n}c_{k}+c_{k+1}\right] \\ 
-\frac{n}{T}\frac{2k_{B}T}{m}\left( \frac{1}{2}D+k\right) \left( \left( -%
\frac{1}{nk_{B}T}T\frac{\partial p^{(0)}}{\partial T}+2k+1\right)
c_{k}-\left( k+1\right) c_{k+1}-kc_{k-1}+\frac{\partial c_{k}}{\partial T}-%
\frac{\partial c_{k+1}}{\partial T}\right) \\ 
\frac{2}{Dk_{B}T}\left( \xi _{1}^{\nabla u}\left[ f_{0}\right]
-p^{(0)c}\right) \left( k\left( c_{k-1}-c_{k}\right) +T\frac{\partial c_{k}}{%
\partial T}\right) \\ 
\frac{2nk_{B}T}{m}\sqrt{\frac{D-1}{D}}\frac{\left( D+k\right) \left(
D+k+2\right) }{4}\left( c_{k+1}-c_{k}\right)%
\end{array}%
\right)
\end{equation}

Finally, it is useful to note that the lowest order coefficients are related
to the kinetic parts of the transport coefficients (see Appendix \ref%
{AppExpandFluxes}). Specifically, the first order contribution to the
pressure tensor takes the usual form 
\begin{equation}
P_{ij}^{(1)}=-\eta \left( \partial _{i}u_{j}+\partial _{j}u_{i}-\frac{2}{D}%
\delta _{ij}\left( \overrightarrow{\nabla }\cdot \overrightarrow{u}\right)
\right) -\gamma \delta _{ij}\left( \overrightarrow{\nabla }\cdot 
\overrightarrow{u}\right)
\end{equation}%
where the shear viscosity, $\eta $, has a kinetic contribution given by%
\begin{equation}
\eta ^{K}=-2nk_{B}T\left( \frac{k_{B}T}{m}\right) \sqrt{\frac{D}{D-1}}%
a_{0}^{\partial u},
\end{equation}%
and the bulk viscosity $\gamma $ has no kinetic contribution. The the first
order contribution to heat flux vector is 
\begin{equation}
\overrightarrow{q}^{\left( 1\right) }\left( \overrightarrow{r},t\right)
=-\mu \overrightarrow{\nabla }\rho -\kappa \overrightarrow{\nabla }T
\end{equation}%
where $\kappa $ is the coefficient of thermal conductivity and $\mu $ is a
new transport coefficient characterizing the way in which density gradients
can cause heat flow due to differential cooling rates. It vanishes in the
elastic limit. The kinetic parts of these transport coefficients are given
by 
\begin{eqnarray}
\mu ^{K} &=&\left( nk_{B}T\left( \frac{k_{B}T}{m}\right) \frac{D+2}{2}%
\right) a_{1}^{\rho } \\
\kappa ^{K} &=&\left( nk_{B}T\left( \frac{k_{B}T}{m}\right) \frac{D+2}{2}%
\right) a_{1}^{T}.  \notag
\end{eqnarray}%
These expressions are exact if $f_{0}$ is replaced by a Gaussian in eq.(\ref%
{ff1}) but if the first order correction is written in terms of $f_{0}$ then
there are terms in $c_{2}$ which would contribute\ (as discussed in the
Appendix) in principle but which would in any case be dropped here since
they are of order $c_{2}a_{s_{0}}^{\left( \gamma \right) }$. The collisional
contributions will be discussed below.

\subsection{Lowest order approximations}

The simplest nontrivial approximation is to keep only the lowest order
nonzero coefficient in each expansion in eq.(\ref{cex}). This means $%
a_{1}^{\left( n\right) },a_{1}^{\left( T\right) },a_{2}^{\left( \nabla
u\right) }$ and $a_{0}^{\left( \partial u\right) }$. Since the transport
coefficients are more interesting than the distribution itself, we write
these equations in terms of the kinetic parts of the transport coefficients
giving%
\begin{equation}
\xi _{0}\left( D+2\right) \frac{1}{m}T\frac{\partial \mu ^{K}}{\partial T}%
+I_{11}^{n}\mu ^{K}-\xi _{0}\left( D+2\right) \frac{1}{m}T\left( \frac{%
\partial \ln n\chi _{0}}{\partial n}\right) \kappa ^{K}=\left( nk_{B}T\left( 
\frac{k_{B}T}{m}\right) \frac{D+2}{2}\right) \left( \Omega _{1}^{n}+\frac{%
k_{B}T}{m}\frac{D\left( D+2\right) }{2}c_{2}\right)  \label{l1}
\end{equation}%
\begin{equation*}
\xi _{0}\left( D+2\right) \frac{1}{m}\left[ T\frac{\partial \kappa ^{K}}{%
\partial T}+\left( T\frac{\partial }{\partial T}\ln \xi _{0}\right) \kappa
^{K}\right] +I_{11}^{T}\kappa ^{K}=mn\left( \frac{k_{B}T}{m}\right) ^{2}%
\frac{D+2}{2}\left( \Omega _{1}^{T}+\frac{n}{T}\frac{2k_{B}T}{m}\frac{1}{4}%
D\left( D+2\right) \left( 1+2c_{2}+\frac{\partial c_{2}}{\partial T}\right)
\right)
\end{equation*}%
\begin{align}
\frac{1}{4}\xi _{0}\frac{D+2}{k_{B}T}\left[ T\frac{\partial }{\partial T}%
a_{2}^{\left( \nabla u\right) }+2a_{2}^{\left( \nabla u\right) }\right]
+I_{22}^{\nabla u}a_{2}^{\left( \nabla u\right) }& =\Omega _{2}^{\nabla u}+%
\frac{D+2}{4k_{B}T}\left( \xi _{1}^{\nabla u}\left[ f_{0}\right]
-p^{(0)c}\right) \left( -2c_{2}+T\frac{\partial c_{2}}{\partial T}\right) 
\notag \\
\xi _{0}\frac{D+2}{m}\left( \frac{2k_{B}T}{m}\right) T\frac{\partial \eta
^{K}}{\partial T}+I_{00}^{\partial u}\eta ^{K}& =mn\left( \frac{k_{B}T}{m}%
\right) ^{2}\left[ n\left( \frac{k_{B}T}{m}\right) D\left( D+2\right) -2%
\sqrt{\frac{D}{D-1}}\Omega _{0}^{\partial u}\right]  \notag
\end{align}

The Boltzmann integrals and the source terms are evaluated in a
straightforward manner and the present evaluations were performed as
described in Appendix \ref{AppEvaluation}, making frequent use of symbolic
manipulation The results can be written as%
\begin{eqnarray}
I_{rs}^{\gamma } &=&I_{rs}^{\gamma E}\left[ 1+\sum_{a}\int_{0}^{\infty
}K_{a}^{\ast }\left( -v\right) e^{-\frac{1}{2}v^{2}}v\left( \Delta
_{a}^{\ast }\left( -v\right) S_{rs}^{\gamma }\left( v\right) +\frac{1}{4}%
vg\left( v,\Delta _{a}^{\ast }\left( -v\right) \right) \right) dv\right]
\label{l2} \\
\Omega _{rs}^{\gamma } &=&\Omega _{rs}^{\gamma E}+\chi n^{2}\sigma ^{D}S_{D}%
\frac{1}{\sqrt{2\pi }}\sum_{a}\int_{0}^{\infty }K_{a}^{\ast }\left(
-v\right) e^{-\frac{1}{2}v^{2}}v\left( T_{rs}^{\gamma }\left( v\right)
+U_{rs}^{\gamma }\left( v\right) c_{2}+V_{rs}^{\gamma }\left( v\right) \frac{%
dc_{2}}{dT}\right) dv  \notag
\end{eqnarray}%
with the elastic contributions%
\begin{eqnarray}
I_{11}^{nE} &=&I_{11}^{TE}=n^{2}\sigma ^{D-1}S_{D}\chi \left( \frac{k_{B}T}{m%
}\right) ^{3/2}\frac{2\left( D-1\right) }{\sqrt{\pi }}  \label{l3} \\
I_{22}^{\nabla uE} &=&n^{2}\sigma ^{D-1}S_{D}\chi \left( \frac{k_{B}T}{m}%
\right) ^{1/2}\frac{\left( D-1\right) }{2\sqrt{\pi }}  \notag \\
I_{00}^{\partial uE} &=&\chi n^{2}\sigma ^{D-1}S_{D}\left( \frac{k_{B}T}{m}%
\right) ^{5/2}\frac{4D}{\sqrt{\pi }}  \notag
\end{eqnarray}%
and the inelastic kernals%
\begin{eqnarray}
S_{11}^{n}\left( v\right) &=&S_{11}^{T}\left( v\right) =-\frac{D+8}{16\left(
D-1\right) }\left( v^{2}-1\right)  \label{l4} \\
S_{22}^{\nabla u}\left( v\right) &=&\frac{1}{64\left( D-1\right) }\left(
v^{6}-9v^{4}+\left( 8D+49\right) \allowbreak v^{2}-37-8D\right)  \notag \\
&&-\frac{1}{64\left( D-1\right) }\left( v^{4}-6\allowbreak v^{2}+3\right)
\Delta ^{\ast }\left( v\right)  \notag \\
S_{00}^{\partial u}\left( v\right) &=&\frac{1}{4D}\left( v^{2}-1\right) 
\notag
\end{eqnarray}%
The elastic contributions to the sources are%
\begin{eqnarray}
\Omega _{1}^{nE} &=&\frac{1}{2}\frac{\partial n^{2}\chi }{\partial n}\sigma
^{D}S_{D}\left( \frac{k_{B}T}{m}\right) \frac{D+5}{4}c_{2}  \label{l6} \\
\Omega _{1}^{TE} &=&-n^{2}\sigma ^{D}\chi S_{D}\frac{k_{B}T}{m}\frac{1}{T}%
\frac{3}{4}\left( 1+2c_{2}+\frac{dc_{2}}{dT}\right)  \notag \\
\Omega _{1}^{\nabla uE} &=&n^{2}\sigma ^{D}\chi S_{D}\frac{D-7}{8D}c_{2} 
\notag \\
\Omega _{1}^{\partial uE} &=&-n^{2}\sigma ^{D}\chi S_{D}\frac{k_{B}T}{m}%
\frac{1}{2}\sqrt{\frac{D-1}{D}}  \notag
\end{eqnarray}%
and the inelastic kernals are%
\begin{eqnarray}
T_{11}^{n}\left( v\right) &=&\frac{1}{2}\frac{\partial \ln n^{2}\chi }{%
\partial n}\left( \frac{k_{B}T}{m}\right) \frac{1}{4}\left( \left(
v^{2}-3\right) g\left( v,\Delta _{a}^{\ast }\left( -v\right) \right)
-2\Delta ^{\ast }\left( v\right) \left( g\left( v,\Delta _{a}^{\ast }\left(
-v\right) \right) +v\right) \right)  \label{l7} \\
T_{11}^{T}\left( v\right) &=&\frac{k_{B}}{16m}\left( 2\left( \left(
v^{2}-1\right) g\left( v,\Delta _{a}^{\ast }\left( -v\right) \right)
+v^{3}+5v\right) \Delta _{a}^{\ast }\left( -v\right) -\left(
9-4v^{2}+v^{4}\right) g\left( v,\Delta _{a}^{\ast }\left( -v\right) \right)
\right)  \notag \\
T_{22}^{\nabla u}\left( v\right) &=&\frac{1}{8D}\left( \left( 2\Delta ^{\ast
}\left( v\right) +3-v^{2}\right) g\left( v,\Delta _{a}^{\ast }\left(
-v\right) \right) +v\Delta _{a}^{\ast }\left( -v\right) \left(
v^{2}-1-\Delta _{a}^{\ast }\left( -v\right) \right) \right)  \notag \\
T_{00}^{\partial u}\left( v\right) &=&\frac{1}{2}\sqrt{\frac{D-1}{D}}\left( 
\frac{k_{B}T}{m}\right) \left( v\Delta _{a}^{\ast }\left( -v\right) -g\left(
v,\Delta _{a}^{\ast }\left( -v\right) \right) \right)  \notag
\end{eqnarray}%
and%
\begin{eqnarray}
U_{11}^{n}\left( v\right) &=&\frac{1}{2}\frac{\partial \ln n^{2}\chi }{%
\partial n}\left( \frac{k_{B}T}{m}\right) \frac{1}{64}\left(
v^{6}-9v^{4}+\left( 49+8D\right) v^{2}-37-8D\right) g\left( v,\Delta
_{a}^{\ast }\left( -v\right) \right) \\
&&+\frac{1}{2}\frac{\partial \ln n^{2}\chi }{\partial n}\left( \frac{k_{B}T}{%
m}\right) \frac{1}{32}\Delta _{a}^{\ast }\left( -v\right) \left(
-v^{4}+6v^{2}-3\right) \left( g\left( v,\Delta _{a}^{\ast }\left( -v\right)
\right) +v\right)  \notag \\
U_{11}^{T}\left( v\right) &=&\frac{k_{B}}{m}\frac{1}{256}\left(
-v^{8}+14v^{6}+\left( -8D-88\right) v^{4}+\left( 126+48D\right)
v^{2}-\allowbreak 24D+33\right) g\left( v,\Delta _{a}^{\ast }\left(
-v\right) \right)  \notag \\
&&+\frac{k_{B}}{m}\frac{1}{128}\Delta _{a}^{\ast }\left( -v\right) \left(
\left( v^{6}-11v^{4}+21v^{2}-3\right) g\left( v,\Delta _{a}^{\ast }\left(
-v\right) \right) +v\left( v^{6}-5v^{4}+9v^{2}-57\right) \right)  \notag \\
U_{22}^{\nabla u}\left( v\right) &=&\frac{1}{128D}v\Delta _{a}^{\ast }\left(
-v\right) \left( \Delta _{a}^{\ast }\left( -v\right) \left(
-v^{4}+10v^{2}-15\right) +v^{6}-11v^{4}+v^{2}\left( 61+8D\right)
-123-24D\right) \allowbreak \allowbreak  \notag \\
&&+\frac{1}{128D}g\left( v,\Delta _{a}^{\ast }\left( -v\right) \right)
\left( 2\Delta _{a}^{\ast }\left( -v\right) \left( v^{4}-6v^{2}+3\right)
-v^{6}+9v^{4}+\left( 8D-65\right) v^{2}-8D+53\right) \allowbreak \\
U_{00}^{\partial u}\left( v\right) &=&\sqrt{\frac{D-1}{D}}\left( \frac{k_{B}T%
}{m}\right) \frac{1}{32}\left( v\Delta _{a}^{\ast }\left(
v^{4}-10v^{2}+15\right) -g\left( v,\Delta _{a}^{\ast }\left( -v\right)
\right) \left( v^{4}-6v^{2}+3\right) \right) .  \notag
\end{eqnarray}%
The only non-vanishing coefficient of the temperature derivative is 
\begin{eqnarray}
V_{11}^{T}\left( v\right) &=&\frac{1}{128}g\left( v,\Delta _{a}^{\ast
}\left( -v\right) \right) \allowbreak \left( -v^{6}+9v^{4}-57v^{2}+45\right)
\\
&&+\frac{1}{64}\Delta _{a}^{\ast }\left( -v\right) \left( \left(
v^{4}-6v^{2}+3\right) g\left( v,\Delta _{a}^{\ast }\left( -v\right) \right)
+\left( v^{4}+6v^{2}-33\right) v\right) .  \notag
\end{eqnarray}

The collisional contributions to the shear and bulk viscosity are, in this
approximation,%
\begin{eqnarray}
\eta ^{C} &=&\frac{2}{3}\theta \eta ^{K}+\frac{D}{D+2}\gamma _{1}  \label{c1}
\\
\gamma &=&\gamma _{1}-\left( nk_{B}T\right) a_{2}^{\nabla u}\frac{%
S_{D}n^{\ast }\chi }{32D\sqrt{2\pi }}\sum_{a}\int_{0}^{\infty }K_{a}^{\ast
}\left( -v\right) e^{-\frac{1}{2}v^{2}}v\left( 3-6v^{2}+v^{4}\right) g\left(
v,\Delta _{a}^{\ast }\left( -v\right) \right) dv  \notag \\
\mu ^{C} &=&\theta \mu ^{K}  \notag \\
\kappa ^{C} &=&\theta \kappa ^{K}+\frac{D}{2}\frac{k_{B}}{m}\gamma _{1} 
\notag \\
&&-m\chi n^{2}\sigma ^{D+1}\left( \frac{k_{B}T}{m}\right) ^{3/2}\frac{1}{4T%
\sqrt{\pi }}\frac{S_{D}}{D}c_{2}\left[ 1+\frac{1}{4}\sum_{a}\int_{0}^{\infty
}K_{a}^{\ast }\left( -v\right) v^{2}e^{-\frac{1}{2}v^{2}}\left(
v^{2}-3\right) g\left( v,\Delta _{a}^{\ast }\left( -v\right) \right) dv%
\right]  \notag
\end{eqnarray}%
with%
\begin{eqnarray}
\gamma _{1} &=&m\sigma n\left( \frac{k_{B}T}{m}\right) ^{\frac{1}{2}}\frac{%
S_{D}n^{\ast }\chi }{D^{2}\sqrt{\pi }}  \label{c2} \\
&&\times \left[ 1-\frac{1}{16}c_{2}+\frac{1}{4}\sum_{a}\int_{0}^{\infty
}K_{a}^{\ast }\left( -v\right) e^{-\frac{1}{2}v^{2}}v^{2}g\left( v,\Delta
_{a}^{\ast }\left( -v\right) \right) \left( 1+\frac{1}{16}c_{2}\left(
v^{4}-10v^{2}+15\right) \right) dv\right]  \notag \\
\theta &=&\frac{3S_{D}}{2D\left( D+2\right) }n^{\ast }\chi \left[ 1+\frac{1}{%
2\sqrt{2\pi }}\sum_{a}\int_{0}^{\infty }K_{a}^{\ast }\left( -v\right) ve^{-%
\frac{1}{2}v^{2}}\left( v^{2}-1\right) g\left( v,\Delta _{a}^{\ast }\left(
-v\right) \right) dv\right]  \notag
\end{eqnarray}%
Finally, the first order corrections to the heat source are%
\begin{eqnarray}
\xi _{0}\left[ f_{1}\right] &=&-\left( \overrightarrow{\nabla }\cdot 
\overrightarrow{u}\right) a_{2}^{\nabla u}n^{2}\sigma ^{D}\chi S_{D}\left( 
\frac{k_{B}T}{m\sigma ^{2}}\right) ^{1/2}\frac{k_{B}T}{32\sqrt{\pi }}%
\sum_{a}\int_{0}^{\infty }K_{a}^{\ast }\left( -v\right) ve^{-\frac{1}{2}%
v^{2}}\Delta _{a}^{\ast }\left( -v\right) \left( v^{4}-6v^{2}+3\right) dv
\label{h1} \\
\xi _{1}\left[ f_{0}\right] &=&\left( \overrightarrow{\nabla }\cdot 
\overrightarrow{u}\right) n^{2}\sigma ^{D}\chi S_{D}\frac{k_{B}T}{2\sqrt{%
2\pi }D}\sum_{a}\int_{0}^{\infty }\;K_{a}^{\ast }\left( -v\right) \Delta
_{a}^{\ast }\left( -v\right) v^{2}e^{-\frac{1}{2}v^{2}}dv  \notag \\
&&+c_{2}\left( \overrightarrow{\nabla }\cdot \overrightarrow{u}\right)
n^{2}\sigma ^{D}\chi S_{D}\frac{k_{B}T}{32D\sqrt{2\pi }}\sum_{a}\int_{0}^{%
\infty }\;K_{a}^{\ast }\left( -v\right) \Delta _{a}^{\ast }\left( -v\right)
e^{-\frac{1}{2}v^{2}}\left( 15-10v^{2}+v^{4}\right) v^{2}dv  \notag
\end{eqnarray}%
Equations (\ref{l1})-(\ref{h1}) are the primary results of this paper. They
give a prescription for the evaluation of the transport properties for an
arbitrary model of energy dissipation at the Navier-Stokes level and in the
usual, lowest Sonine approximation. In the next Subsection, these are
illustrated by using them to give the transport properties of a simple
granular fluid.

\subsection{Transport in simple granular fluids}

For the simple granular fluid, recall that $\Delta ^{\ast }\left( v\right)
=\left( 1-\alpha ^{2}\right) \frac{1}{2}v^{2}$ and $g\left( v,\Delta \right)
=v\left( \alpha -1\right) $. Since there is no other energy scale, the
coefficients of the first order solution must scale with temperature as 
\begin{eqnarray}
a_{1}^{\left( n\right) } &\sim &T^{-1/2} \\
a_{1}^{\left( T\right) } &\sim &T^{-3/2}  \notag \\
a_{2}^{\left( \nabla u\right) } &\sim &T^{-1/2}  \notag \\
a_{0}^{\left( \partial u\right) } &\sim &T^{-3/2}  \notag \\
\xi _{0} &\sim &T^{3/2}  \notag
\end{eqnarray}%
giving%
\begin{gather}
\left[ \xi _{0}\left( D+2\right) \frac{3}{2m}+I_{11}^{n}\right] \mu ^{K}+\xi
_{0}\left( D+2\right) \frac{1}{m}T\left( \frac{\partial \ln n\chi _{0}}{%
\partial n}\right) \kappa ^{K}=\left( nk_{B}T\left( \frac{k_{B}T}{m}\right) 
\frac{D+2}{2}\right) \left( \Omega _{1}^{n}+\frac{D\left( D+2\right) }{2}%
\frac{k_{B}T}{m}c_{2}\right) \\
\left[ \xi _{0}\left( D+2\right) \frac{2}{m}+I_{11}^{T}\right] \kappa
^{K}=mn\left( \frac{k_{B}T}{m}\right) ^{2}\frac{D+2}{2}\Omega _{1}^{T}+\frac{%
1}{T}mn^{2}\left( \frac{k_{B}T}{m}\right) ^{3}\frac{D\left( D+2\right) ^{2}}{%
4}\left( 1+2c_{2}\right)  \notag \\
\left( \frac{3}{8}\xi _{0}\frac{D+2}{k_{B}T}+I_{22}^{\nabla u}\right)
a_{2}^{\left( \nabla u\right) }=\Omega _{2}^{\nabla u}-\frac{D+2}{2k_{B}T}%
\left( \xi _{1}^{\nabla u}\left[ f_{0}\right] -p^{(0)c}\right) c_{2}  \notag
\\
\left( \xi _{0}\frac{D+2}{m}\left( \frac{k_{B}T}{m}\right) +I_{00}^{\partial
u}\right) \eta ^{K}=mn\left( \frac{k_{B}T}{m}\right) ^{2}\left[ n\left( 
\frac{k_{B}T}{m}\right) D\left( D+2\right) -2\sqrt{\frac{D}{D-1}}\Omega
_{0}^{\partial u}\right] .  \notag
\end{gather}%
From the zeroth order solution, one has 
\begin{equation}
\xi _{0}\left[ f_{0}\right] =-\left( 1-\alpha ^{2}\right) n^{\ast }\chi 
\frac{S_{D}}{2\sqrt{\pi }}\left( \frac{k_{B}T}{m\sigma ^{2}}\right)
^{1/2}nk_{B}T
\end{equation}%
Equations (\ref{l2}) are easily evaluated giving the Boltzmann integrals 
\begin{eqnarray}
I_{11}^{n} &=&I_{11}^{T}=-n^{2}\sigma ^{D-1}S_{D}\chi \left( \frac{k_{B}T}{m}%
\right) ^{3/2}\frac{1}{8\sqrt{\pi }}\left( \alpha +1\right) \left( 3\alpha
\left( D+8\right) -11D-16\right) \\
I_{22}^{\nabla u} &=&-n^{2}\sigma ^{D-1}S_{D}\chi \left( \frac{k_{B}T}{m}%
\right) ^{1/2}\frac{1}{128\sqrt{\pi }}\left( \alpha +1\right) \left(
30\alpha ^{3}-30\alpha ^{2}+105\alpha +24\alpha D-56D-73\right)  \notag \\
I_{00}^{\partial u} &=&n^{2}\sigma ^{D-1}S_{D}\chi \left( \frac{k_{B}T}{m}%
\right) ^{5/2}\frac{1}{\sqrt{\pi }}\left( 1+\alpha \right) \left( 3-3\alpha
+2D\right)  \notag
\end{eqnarray}%
and sources%
\begin{eqnarray}
\Omega _{1}^{n} &=&\frac{1}{2}\frac{\partial n^{2}\chi }{\partial n}\sigma
^{D}S_{D}\left( \frac{k_{B}T}{m}\right) \frac{1}{16}\left( 1+\alpha \right) %
\left[ 6\alpha \left( 1-\alpha \right) -\left( 3\alpha ^{2}-3\alpha
+10+2D\right) c_{2}\right] \\
\Omega _{1}^{T} &=&\frac{3}{16}n^{2}\sigma ^{D}\chi S_{D}\frac{k_{B}}{m}%
\left( \alpha +1\right) ^{2}\left[ 2\alpha -1+\left( \alpha +1\right) c_{2}%
\right]  \notag \\
\Omega _{2}^{\nabla u} &=&\frac{3}{64D}n^{2}\sigma ^{D}\chi S_{D}\left(
1+\alpha \right) \left( \left( 5\alpha -1\right) \left( 1-\alpha \right)
\left( 1+\alpha \right) -\frac{1}{6}c_{2}\left( 15\alpha ^{3}-3\alpha
^{2}+3\left( 4D+15\right) \alpha -\left( 20D+1\right) \right) \right)  \notag
\\
\Omega _{0}^{\partial u} &=&-\frac{1}{8}n^{2}\sigma ^{D}\chi S_{D}k_{B}T%
\sqrt{\left( \frac{D-1}{D}\right) }\left( \alpha +1\right) \frac{3\alpha -1}{%
m}  \notag
\end{eqnarray}%
The low density (Boltzmann)\ transport coefficients in the elastic ($\alpha
=1$) limit can be read off and are%
\begin{eqnarray}
\kappa _{0} &=&k_{B}\left( \frac{k_{B}T}{m}\right) ^{1/2}\frac{\sqrt{\pi }}{%
8\sigma ^{D-1}S_{D}}\frac{D\left( D+2\right) ^{2}}{D-1} \\
\eta _{0} &=&\sqrt{\frac{mk_{B}T}{\pi }}\frac{\left( D+2\right) \pi }{4S_{D}}%
\sigma ^{1-D}  \notag \\
\mu _{0} &=&\gamma _{0}=0.  \notag
\end{eqnarray}%
To facilitate comparison of the finite density transport coefficients to
those of ref.\cite{DuftyGranularTransport}, it is useful to introduce
dimensionless quantities%
\begin{eqnarray}
\nu _{T}^{\ast } &=&\chi \frac{D-1}{2D}\left( \alpha +1\right) \left( 1+%
\frac{3\left( D+8\right) \left( 1-\alpha \right) }{8\left( D-1\right) }%
\right) \\
\nu _{\eta }^{\ast } &=&\chi \left( 1-\frac{1}{4D}\left( 1-\alpha \right)
\left( 2D-3\alpha -3\right) \right)  \notag \\
\nu _{\gamma }^{\ast } &=&-\frac{1}{48}\chi \left( \frac{\alpha +1}{2}%
\right) \left( 30\alpha ^{3}-30\alpha ^{2}+105\alpha +24\alpha
D-56D-73\right)  \notag \\
\zeta ^{\ast } &=&\frac{D+2}{4D}\chi \left( 1-\alpha ^{2}\right)  \notag
\end{eqnarray}%
so that

\begin{eqnarray}
I_{11}^{n} &=&I_{11}^{T}=\frac{2D}{\sqrt{\pi }}S_{D}nn^{\ast }\left( \frac{%
k_{B}T}{m}\right) \left( \frac{k_{B}T}{m\sigma ^{2}}\right) ^{1/2}\nu
_{T}^{\ast } \\
I_{00}^{\partial u} &=&\frac{4D}{\sqrt{\pi }}S_{D}nn^{\ast }\left( \frac{%
k_{B}T}{m}\right) ^{2}\left( \frac{k_{B}T}{m\sigma ^{2}}\right) ^{1/2}\nu
_{\eta }^{\ast }  \notag \\
I_{22}^{\nabla u} &=&\frac{3}{4\sqrt{\pi }}S_{D}nn^{\ast }\left( \frac{k_{B}T%
}{m\sigma ^{2}}\right) ^{1/2}\nu _{\gamma }^{\ast }  \notag \\
\frac{1}{m}\xi ^{(0)} &=&-\frac{2D}{\sqrt{\pi }\left( D+2\right) }%
S_{D}nn^{\ast }\frac{k_{B}T}{m}\left( \frac{k_{B}T}{m\sigma ^{2}}\right)
^{1/2}\zeta ^{\ast }  \notag
\end{eqnarray}%
and then%
\begin{eqnarray}
\kappa ^{K} &=&\kappa _{0}\frac{D-1}{D}\left[ \nu _{T}^{\ast }-2\zeta ^{\ast
}\right] ^{-1}\left( 1+2c_{2}+\frac{3}{8D\left( D+2\right) }n^{\ast }\chi
S_{D}\left( \alpha +1\right) ^{2}\left( 2\alpha -1+\left( \alpha +1\right)
c_{2}\right) \right) \\
\mu ^{K} &=&2\kappa _{0}\frac{T}{n}\left[ \nu _{T}^{\ast }-2\zeta ^{\ast }%
\right] ^{-1}  \notag \\
&&\times \left( \kappa ^{K\ast }\zeta ^{\ast }+\frac{D-1}{D}c_{2}+\frac{1}{2}%
\frac{\partial n^{\ast 2}\chi }{\partial n^{\ast }}S_{D}\frac{3\left(
D-1\right) }{8D^{2}\left( D+2\right) }\left( 1+\alpha \right) \left(
-2\alpha \left( 1-\alpha \right) +\frac{1}{6}\left( 3\alpha ^{2}-3\alpha
+10+2D\right) 2c_{2}\right) \right)  \notag \\
\eta ^{K} &=&\eta _{0}\left[ \nu _{\eta }^{\ast }-\frac{1}{2}\zeta ^{\ast }%
\right] ^{-1}\left( 1+\frac{S_{D}}{4D\left( D+2\right) }n^{\ast }\chi \left(
\alpha +1\right) \left( 3\alpha -1\right) \right)  \notag \\
a_{2}^{\nabla u} &=&\frac{\eta _{0}}{nk_{B}T}\frac{1}{4D\left( D+2\right) }%
\left[ \nu _{\gamma }^{\ast }-D\zeta ^{\ast }\right] ^{-1}\left( \chi
n^{\ast }S_{D}\lambda ^{\ast }+\frac{16D\left( D+2\right) }{nk_{B}T}%
p^{(0)c}\left( \frac{1}{3}-\alpha \right) c_{2}\right)  \notag
\end{eqnarray}%
where $\kappa ^{K\ast }=\kappa ^{K}/\kappa _{0}$ and 
\begin{equation}
\lambda ^{\ast }=\left( 1+\alpha \right) \left[ \left( 5\alpha -1\right)
\left( 1-\alpha \right) \left( 1+\alpha \right) -\frac{1}{6}c_{2}\left(
15\alpha ^{3}-3\alpha ^{2}+3\left( 4D+15\right) \alpha -\left( 20D+1\right)
\right) \right]
\end{equation}

The collisional parts of the transport coefficients are%
\begin{eqnarray}
\eta ^{C} &=&\eta ^{K}\frac{S_{D}}{D\left( D+2\right) }n^{\ast }\chi \left( 
\frac{1+\alpha }{2}\right) +\frac{D}{D+2}\gamma \\
\gamma &=&\eta _{0}\frac{4S_{D}^{2}}{\pi \left( D+2\right) D^{2}}\left( 
\frac{1+\alpha }{2}\right) n^{\ast 2}\chi \left( 1-\frac{1}{16}c_{2}\right) 
\notag \\
\mu ^{C} &=&\frac{3S_{D}}{2D\left( D+2\right) }n^{\ast }\chi \left( \frac{%
1+\alpha }{2}\right) \mu ^{K}  \notag \\
\kappa ^{C} &=&\frac{3S_{D}}{2D\left( D+2\right) }n^{\ast }\chi \left( \frac{%
1+\alpha }{2}\right) \kappa ^{K}+\kappa _{0}\left( 1+\alpha \right) \frac{%
2S_{D}^{2}\left( D-1\right) }{\pi D^{2}\left( D+2\right) ^{2}}\allowbreak
n^{\ast 2}\chi \left( 1-\frac{7}{16}c_{2}\right)  \notag
\end{eqnarray}%
and the first order correction to the cooling rate is%
\begin{eqnarray}
\xi ^{(1)} &=&\xi _{0}\left[ f_{1}\right] +\xi _{1}\left[ f_{0}\right] \\
\xi _{0}\left[ f_{1}\right] &=&-a_{2}^{\nabla u}\left( \overrightarrow{%
\nabla }\cdot \overrightarrow{u}\right) \left( \frac{3}{2}nk_{B}T\right) 
\frac{nk_{B}T}{\eta _{0}}\frac{D+2}{64}\left( 1-\alpha ^{2}\right) \chi 
\notag \\
\xi _{1}\left[ f_{0}\right] &=&\left( \overrightarrow{\nabla }\cdot 
\overrightarrow{u}\right) \left( \frac{3}{2}nk_{B}T\right) \frac{S_{D}}{4D}%
n^{\ast }\chi \left( 1-\alpha ^{2}\right)  \notag
\end{eqnarray}%
Taking into account that the quantities $\zeta =-\frac{2}{3nk_{B}T}\xi ,$ $%
c^{\ast }=2c_{2}$ and $c_{D}=\frac{1}{2}a_{2}^{\nabla u}$ are used in ref.%
\cite{DuftyGranularTransport}, it is easy to verify that the present
expressions agree for the special case of $D=3$ with those of ref.\cite%
{DuftyGranularTransport} up to terms of order $c^{\ast }a_{s_{0}}^{\left(
\gamma \right) }$ except for the expression for $a_{2}^{\nabla u}$. The
expressions given for this quantity in ref.\cite{DuftyGranularTransport} are
incorrect and according to \cite{priv:2005} the correct expressions agree
with those given here except for the terms $-\frac{1}{6}c_{2}\left( 3\left(
4D+15\right) \alpha -\left( 20D+1\right) \right) $ in the expression for $%
\lambda ^{\ast }$ which give $-\frac{1}{6}c_{2}\left( 81\alpha -61\right) $
for $D=3$ whereas ref\cite{priv:2005} gives $\frac{1}{12}c^{\ast }\left(
159\alpha -19\right) $.

\section{Conclusions}

A normal solution to the Enskog approximation for a hard-sphere gas with
energy loss has been determined using the Chapman-Enskog procedure to first
order in the gradients, and the transport properties given to second order
thus specifying the Navier-Stokes hydrodynamic description. The zeroth order
distribution function, which describes the homogeneous cooling state, was
expanded about equilibrium, eq.(\ref{hsc-distribution}), and the equations
for the coefficients given. The required collision integrals were expressed
in terms of a generating function allowing for evaluation using symbolic
mathematical packages and the explicit form of the required integrals needed
to determine the first correction to the Gaussian approximation were given
explicitly in the form of one-dimensional integrals. The expressions for the
transport properties have similarly been reduced to simple quadratures for
the standard lowest-Sonine approximation. The Navier-Stokes equations for
such a system thus take the usual form, eq.(\ref{balance}), with pressure
tensor%
\begin{equation}
P_{ij}=p^{(0)}\delta _{ij}-\eta \left( \partial _{i}u_{j}+\partial _{j}u_{i}-%
\frac{2}{D}\delta _{ij}\left( \overrightarrow{\nabla }\cdot \overrightarrow{u%
}\right) \right) -\gamma \delta _{ij}\left( \overrightarrow{\nabla }\cdot 
\overrightarrow{u}\right)
\end{equation}%
and heat-flux vector%
\begin{equation}
\overrightarrow{q}\left( \overrightarrow{r},t\right) =-\mu \overrightarrow{%
\nabla }\rho -\kappa \overrightarrow{\nabla }T
\end{equation}%
where $\mu $ represents a new transport coefficient not present when the
collision conserve energy. Equations (\ref{l1})-(\ref{l7}) determine the
kinetic parts of the transport coefficients and eq.s(\ref{c1}) and (\ref{c2}%
) determine their collisional parts. The pressure, $p^{(0)}$, is given in
eq.(\ref{pressure}) and the source term in the temperature equation, which
accounts for the cooling, is given by eq.(\ref{s1}) and eq.(\ref{h1}).
Finally, as a simple application, the transport properties for a granular
fluid in $D$ dimensions were given and previous results for the special case 
$D=3$ recovered.

\begin{figure*}[tbp]
\includegraphics[angle=0,scale=0.4]{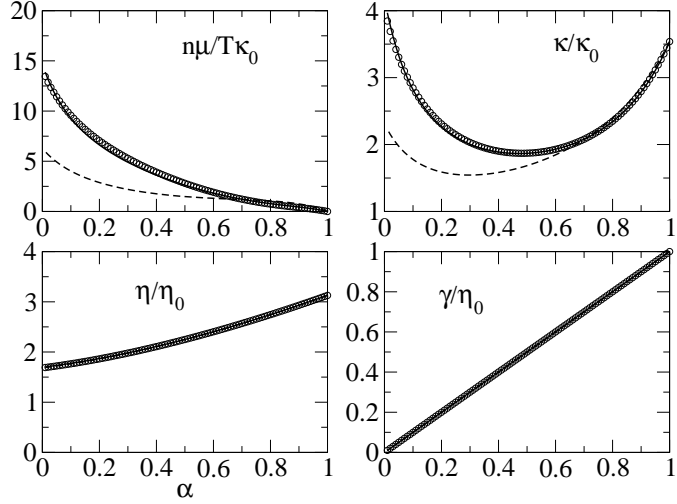}
\caption{The four transport coefficients for a simple granular fluid in 3
dimensions at reduced density $n\protect\sigma^{3}=0.5$. The lines are the
results of ref.\protect\cite{DuftyGranularTransport}, the circles are the
results of this paper and the broken line is the Gaussian approximation.}
\label{fig1}
\end{figure*}

One question which has been left unanswered is whether we can judge the
effect of neglecting some of the contributions to the transport coefficients
due to the non-Gaussian part of the zeroth order distribution. To answer
this, we show in figure 1 the four transport coefficients for a simple
granular fluid in three dimensions at a moderately high density of $n\sigma
^{3}=0.5$ . The values are calculated based on the expressions of ref.\cite%
{DuftyGranularTransport} which include all contributions in $c_{2}$, those
given in the previous Section and the results of the Gaussian approximation, 
$c_{2}=0$. All three approximations are in agreement for the shear and bulk
viscosity, but the Gaussian approximation gives quantitatively poor results
for thermal conductivity and the new transport coefficient. On the other
hand, the expressions given above are in good agreement with the full
expressions for the entire range of the coefficient of restitution thus
giving some justification for the approximations used here.

\begin{acknowledgements}
This work was supported in part by the European Space Agency
under contract number C90105.
\end{acknowledgements}

\appendix

\section{Expansion of the collision operator}

\label{AppExpandOperator}

The collision operator is%
\begin{equation}
J\left[ f,f\right] =-\int dx_{2}\;\overline{T}_{-}\left( 12\right) \chi
\left( \overrightarrow{q}_{1},\overrightarrow{q}_{2};\left[ n\right] \right)
f\left( \overrightarrow{q}_{1},\overrightarrow{v}_{1};t\right) f\left( 
\overrightarrow{q}_{2},\overrightarrow{v}_{2};t\right)
\end{equation}%
where $\chi \left( \overrightarrow{q}_{1},\overrightarrow{q}_{2};\left[ n%
\right] \right) $ is the local equilibrium pair distribution function which
is in general a functional of the local density and we write binary
collision operator $\overline{T}_{-}\left( 12\right) $ as 
\begin{eqnarray}
\overline{T}_{-}\left( 12\right) &=&\delta \left( q_{12}-\sigma \right) 
\overline{T}_{-}^{\prime }\left( 12\right) \\
\overline{T}_{-}^{\prime }\left( 12\right) &=&\left[ \sum_{a}J_{a}\left( 
\overrightarrow{v}_{1},\overrightarrow{v}_{2}\right) \left( \widehat{b}%
_{a}\right) ^{-1}K_{a}\left( \widehat{q}_{12}\cdot \overrightarrow{v}%
_{12}\right) -1\right] \Theta \left( -\overrightarrow{v}_{12}\cdot \widehat{q%
}_{12}\right) \delta \left( q_{12}-\sigma \right) \overrightarrow{v}%
_{12}\cdot \widehat{q}_{12}.  \notag
\end{eqnarray}%
Integrating over the argument of the delta-function gives%
\begin{equation}
J\left[ f,f\right] =-\sigma ^{D-1}\int d\overrightarrow{v}_{2}d\widehat{q}%
_{12}\;\overline{T}_{-}^{\prime }\left( 12\right) \chi \left( 
\overrightarrow{q}_{1},\overrightarrow{q}_{1}-\sigma \widehat{q}_{12};\left[
n\right] \right) f\left( \overrightarrow{q}_{1},\overrightarrow{v}%
_{1};t\right) f\left( \overrightarrow{q}_{1}-\sigma \widehat{q}_{12},%
\overrightarrow{v}_{2};t\right) .
\end{equation}%
A gradient expansion of the nonlocal terms requires first an expansion of
the one-body distribution%
\begin{equation}
f\left( \overrightarrow{q}_{1}-\sigma \widehat{q}_{12},\overrightarrow{v}%
_{2};t\right) =f\left( \overrightarrow{q}_{1},\overrightarrow{v}%
_{2};t\right) -\sigma \widehat{q}_{12}\cdot \overrightarrow{\nabla }%
_{1}f\left( \overrightarrow{q}_{1},\overrightarrow{v}_{2};t\right) +...
\end{equation}%
For a normal solution, this becomes%
\begin{equation}
f\left( \overrightarrow{q}_{1}-\sigma \widehat{q}_{12},\overrightarrow{v}%
_{2};t\right) =f\left( \overrightarrow{q}_{1},\overrightarrow{v}%
_{2};t\right) -\sigma \widehat{q}_{12}\cdot \sum_{i}\left( \overrightarrow{%
\nabla }_{1}\psi _{i}\left( \overrightarrow{q}_{1}\right) \right) \frac{%
\delta }{\delta \psi _{i}\left( \overrightarrow{q}_{1}\right) }f\left( 
\overrightarrow{q}_{1},\overrightarrow{v}_{2};t\right) +...
\end{equation}%
and of the nonlocal dependence on the density of the pair distribution
function%
\begin{eqnarray}
&&\chi \left( \overrightarrow{q}_{1},\overrightarrow{q}_{1}-\sigma \widehat{q%
}_{12};\left[ n\right] \right) \\
&=&\chi _{0}\left( \sigma ;n\left( \overrightarrow{q}_{1}\right) \right)
+\int d\overrightarrow{r}\;\left( n\left( \overrightarrow{r}\right) -n\left( 
\overrightarrow{q}_{1}\right) \right) \left[ \frac{\delta }{\delta n\left( 
\overrightarrow{r}\right) }\chi \left( \overrightarrow{q}_{1},%
\overrightarrow{q}_{1}-\sigma \widehat{q}_{12};\left[ n\right] \right) %
\right] _{n\left( \overrightarrow{q}_{1}\right) }+...  \notag \\
&=&\chi _{0}\left( \sigma ;n\left( \overrightarrow{q}_{1}\right) \right)
+\left( \overrightarrow{\nabla }_{1}n\left( \overrightarrow{q}_{1}\right)
\right) \int d\overrightarrow{r}\;\left( \overrightarrow{r}-\overrightarrow{q%
}_{1}\right) \left[ \frac{\delta }{\delta n\left( \overrightarrow{r}\right) }%
\chi \left( \overrightarrow{q}_{1},\overrightarrow{q}_{1}-\sigma \widehat{q}%
_{12};\left[ n\right] \right) \right] _{n\left( \overrightarrow{q}%
_{1}\right) }+...  \notag
\end{eqnarray}%
which is accurate up to first order in the gradients. For a single-component
system, it can be shown\cite{RET} that the second term reduces to a simple
derivative giving%
\begin{equation}
\chi \left( \overrightarrow{q}_{1},\overrightarrow{q}_{1}-\sigma \widehat{q}%
_{12};\left[ n\right] \right) =\chi _{0}\left( \sigma ;n\left( 
\overrightarrow{q}_{1}\right) \right) -\frac{1}{2}\left( \sigma \widehat{q}%
_{12}\cdot \overrightarrow{\nabla }_{1}n\left( \overrightarrow{q}_{1}\right)
\right) \left. \frac{\partial \chi _{0}\left( \sigma ;n\right) }{\partial n}%
\right| _{n\left( \overrightarrow{q}_{1}\right) }+...
\end{equation}

Using these results, we can write 
\begin{equation}
J\left[ f,f\right] =J_{0}\left[ f,f\right] +J_{1}\left[ f,f\right] +...
\end{equation}%
with%
\begin{equation}
J_{0}\left[ f,f\right] =-\chi _{0}\left( \sigma ;n\left( \overrightarrow{q}%
_{1}\right) \right) \int dx_{2}\;\overline{T}_{-}\left( 12\right) f\left( 
\overrightarrow{q}_{1},\overrightarrow{v}_{1};t\right) f\left( 
\overrightarrow{q}_{1},\overrightarrow{v}_{2};t\right)
\end{equation}%
which is, aside from the prefactor of $\chi _{0}\left( \sigma ;n\left( 
\overrightarrow{q}_{1}\right) \right) $, the Boltzmann collision operator.
The second order term is%
\begin{eqnarray}
J_{1}\left[ f,f\right] &=&\sum_{i}\left( \overrightarrow{\nabla }_{1}\psi
_{i}\left( \overrightarrow{q}_{1}\right) \right) \cdot \sigma ^{D}\chi
_{0}\left( \sigma ;n\left( \overrightarrow{q}_{1}\right) \right) \int d%
\overrightarrow{v}_{2}d\widehat{q}_{12}\;\widehat{q}_{12}\overline{T}%
_{-}^{\prime }\left( 12\right) f\left( \overrightarrow{q}_{1},%
\overrightarrow{v}_{1};t\right) \frac{\delta }{\delta \psi _{i}\left( 
\overrightarrow{q}_{1}\right) }f\left( \overrightarrow{q}_{1},%
\overrightarrow{v}_{2};t\right) \\
&&+\left( \overrightarrow{\nabla }_{1}n\left( \overrightarrow{q}_{1}\right)
\right) \frac{1}{2}\sigma ^{D}\frac{\partial \chi _{0}\left( \sigma ;n\left( 
\overrightarrow{q}_{1}\right) \right) }{\partial n\left( \overrightarrow{q}%
_{1}\right) }\int d\overrightarrow{v}_{2}d\widehat{q}_{12}\;\widehat{q}_{12}%
\overline{T}_{-}^{\prime }\left( 12\right) f\left( \overrightarrow{q}_{1},%
\overrightarrow{v}_{1};t\right) f\left( \overrightarrow{q}_{1},%
\overrightarrow{v}_{2};t\right)  \notag
\end{eqnarray}%
which we write more compactly as%
\begin{equation}
J_{1}\left[ f,f\right] =\sum_{i}\left( \overrightarrow{\nabla }_{1}\psi
_{i}\left( \overrightarrow{q}_{1}\right) \right) \left( \overrightarrow{J}%
_{1}\left[ f,\frac{\delta }{\delta \psi _{i}\left( \overrightarrow{q}%
_{1}\right) }f\right] +\delta _{in}\frac{\partial \ln \chi _{0}\left( \sigma
;n\left( \overrightarrow{q}_{1}\right) \right) }{\partial n\left( 
\overrightarrow{q}_{1}\right) }\overrightarrow{J}_{1}\left[ f,f\right]
\right)
\end{equation}%
with%
\begin{equation}
\overrightarrow{J}_{1}\left[ f,g\right] =\int d\overrightarrow{v}_{2}d%
\widehat{q}_{12}\;\widehat{q}_{12}\overline{T}_{-}^{\prime }\left( 12\right)
f\left( \overrightarrow{q}_{1},\overrightarrow{v}_{1};t\right) g\left( 
\overrightarrow{q}_{1},\overrightarrow{v}_{2};t\right) .
\end{equation}

\section{Expansion of the fluxes and sources}

\label{AppExpandFluxes} The expansion of the fluxes and sources is very
similar to that of the collision operator described in the previous Appendix
so only a few details will be given here.

\subsection{Pressure tensor}

The exact expression for the pressure tensor is $\overleftrightarrow{P}=%
\overleftrightarrow{P}^{K}+\overleftrightarrow{P}^{C}$ with 
\begin{equation}
\overleftrightarrow{P}^{K}\left( \overrightarrow{r},t|f\right) =m\int d%
\overrightarrow{v}_{1}\;f\left( \overrightarrow{r},\overrightarrow{v}%
_{1},t\right) \overrightarrow{V}_{1}\overrightarrow{V}_{1},
\end{equation}%
and the collisional contribution which can be written as%
\begin{eqnarray}
\overleftrightarrow{P}^{C}\left( \overrightarrow{r},t|f\right) &=&-\frac{m}{4%
}\sigma ^{D}\sum_{a}\int d\overrightarrow{v}_{1}d\overrightarrow{v}_{2}d%
\widehat{q}\;\widehat{q}\widehat{q}\left( \widehat{q}\cdot \overrightarrow{v}%
_{12}\right) \Theta \left( -\widehat{q}\cdot \overrightarrow{v}_{12}\right)
\\
&&\times K_{a}\left( \widehat{q}\cdot \overrightarrow{v}_{12}\right)
\int_{0}^{1}dy\;\chi \left( \overrightarrow{r}+\left( 1-y\right) \sigma 
\widehat{q},\overrightarrow{r}-y\sigma \widehat{q};\left[ n\right] \right)
f\left( \overrightarrow{r}+\left( 1-y\right) \sigma \widehat{q},%
\overrightarrow{v}_{1};t\right) f\left( \overrightarrow{r}-y\sigma \widehat{q%
},\overrightarrow{v}_{2};t\right)  \notag \\
&&\times \left( -\overrightarrow{v}_{12}\cdot \widehat{q}-sgn\left( 
\overrightarrow{v}_{12}\cdot \widehat{q}\right) \sqrt{\left( \overrightarrow{%
v}_{12}\cdot \widehat{q}\right) ^{2}-\frac{4}{m}\Delta _{a}\left( \widehat{q}%
\cdot \overrightarrow{v}_{12}\right) }\right) .  \notag
\end{eqnarray}%
Clearly, the expansion of kinetic part is simply due to the expansion of the
distribution function itself%
\begin{equation}
\overleftrightarrow{P}^{K}\left( \overrightarrow{r},t|f\right)
=\sum_{i=0}\epsilon ^{i}\overleftrightarrow{P}^{K(i)}\left( \overrightarrow{r%
},t\right)
\end{equation}%
with%
\begin{equation*}
\overleftrightarrow{P}^{K(i)}\left( \overrightarrow{r},t\right) =%
\overleftrightarrow{P}^{K}\left( \overrightarrow{r},t|f_{i}\right) .
\end{equation*}%
The zeroth order contribution is based on $f_{0}$ which is homogeneous so 
\begin{equation}
P_{ij}^{K(0)}\left( \overrightarrow{r},t|f\right) =m\frac{1}{D}\delta
_{ij}\int d\overrightarrow{V}\;f_{0}\left( \overrightarrow{V}|\psi
_{t}\right) V^{2}=nk_{B}T\delta _{ij}
\end{equation}%
from the definition of the temperature. The first order contribution is%
\begin{equation}
P_{ij}^{K(1)}\left( \overrightarrow{r},t|f\right) =m\int d\overrightarrow{V}%
\;f_{1}\left( \overrightarrow{V}|\psi _{t}\right) V_{i}V_{j}
\end{equation}%
and comparison to the definition of the first order term eq.(\ref{ff1})
shows that the only contribution is due to the shear term%
\begin{eqnarray}
P_{ij}^{K(1)}\left( \overrightarrow{r},t|f\right) &=&m\int d\overrightarrow{V%
}\;f_{0}\left( x_{1}\right) \left[ \sqrt{\frac{D}{D-1}}A^{\left( \partial
u\right) }\left( \overrightarrow{V}\right) \left( V_{l}V_{k}-\frac{1}{D}%
\delta _{lk}V^{2}\right) \left( \partial _{l}u_{k}+\partial _{l}u_{k}-\frac{2%
}{D}\delta _{lk}\overrightarrow{\nabla }\cdot \overrightarrow{u}\right) %
\right] V_{i}V_{j} \\
&=&m\sqrt{\frac{D}{D-1}}\left( \partial _{l}u_{k}+\partial _{l}u_{k}-\frac{2%
}{D}\delta _{lk}\overrightarrow{\nabla }\cdot \overrightarrow{u}\right) 
\frac{1}{D(D+2)}\left( \delta _{il}\delta _{jk}+\delta _{ik}\delta
_{jl}\right) \int d\overrightarrow{V}\;f_{0}\left( x_{1}\right) A^{\left(
\partial u\right) }\left( \overrightarrow{V}\right) V^{4}  \notag
\end{eqnarray}%
and using the expansion of $A^{\left( \partial u\right) }$ in associated
Laguerre polynomials and their orthogonality relation gives%
\begin{eqnarray}
P_{ij}^{K(1)}\left( \overrightarrow{r},t|f\right) &=&mnS_{D}\sqrt{\frac{D}{%
D-1}}\left( \partial _{i}u_{j}+\partial _{j}u_{i}-\frac{2}{D}\delta _{ij}%
\overrightarrow{\nabla }\cdot \overrightarrow{u}\right) \frac{1}{D(D+2)} \\
&&\times \sum_{ij}a_{i}^{\partial u}c_{j}\left( \frac{2k_{B}T}{m}\right)
^{2}\pi ^{-D/2}\int_{0}^{\infty }\exp \left( -x\right) L_{i}^{\frac{D+2}{2}%
}\left( x\right) L_{j}^{\frac{D-2}{2}}\left( x\right) x^{\frac{D+2}{2}}dx 
\notag \\
&=&mnS_{D}\pi ^{-D/2}\sqrt{\frac{D}{D-1}}\left( \partial _{i}u_{j}+\partial
_{j}u_{i}-\frac{2}{D}\delta _{ij}\overrightarrow{\nabla }\cdot 
\overrightarrow{u}\right) \frac{1}{D(D+2)}  \notag \\
&&\times \sum_{ij}\frac{\Gamma \left( \frac{D}{2}+i+2\right) }{\Gamma \left(
i+1\right) }a_{i}^{\partial u}\left( \frac{2k_{B}T}{m}\right) ^{2}\left(
c_{i}-2c_{i+1}+c_{i+2}\right)  \notag \\
&=&nk_{B}T\left[ 2a_{0}^{\partial u}\left( \frac{k_{B}T}{m}\right) \left(
1+c_{2}\right) \sqrt{\frac{D}{D-1}}\right] \left( \partial
_{i}u_{j}+\partial _{j}u_{i}-\frac{2}{D}\delta _{ij}\overrightarrow{\nabla }%
\cdot \overrightarrow{u}\right)  \notag
\end{eqnarray}%
and in the present approximation, we drop the term $c_{2}$.

The collisional part of the stress tensor will have terms arising from the
expansion of the distribution as well as gradient terms arising from its
nonlocality. The former gives a first order contribution of 
\begin{eqnarray}
\overleftrightarrow{P}^{C(11)}\left( \overrightarrow{r},t|f\right) &=&-\frac{%
m}{4}\sigma ^{D}\chi _{0}\left( \sigma ;n\left( \overrightarrow{r}\right)
\right) \sum_{a}\int d\overrightarrow{v}_{1}d\overrightarrow{v}_{2}d\widehat{%
q}\;\widehat{q}\widehat{q}\left( \widehat{q}\cdot \overrightarrow{v}%
_{12}\right) \Theta \left( -\widehat{q}\cdot \overrightarrow{v}_{12}\right)
\\
&&\times K_{a}\left( \widehat{q}_{12}\cdot \overrightarrow{v}_{12}\right) %
\left[ f_{0}\left( \overrightarrow{r},\overrightarrow{v}_{1};t\right)
f_{1}\left( \overrightarrow{r},\overrightarrow{v}_{2};t\right) +f_{1}\left( 
\overrightarrow{r},\overrightarrow{v}_{1};t\right) f_{0}\left( 
\overrightarrow{r},\overrightarrow{v}_{2};t\right) \right]  \notag \\
&&\times \left( -\overrightarrow{v}_{12}\cdot \widehat{q}_{12}-sgn\left( 
\overrightarrow{v}_{12}\cdot \widehat{q}_{12}\right) \sqrt{\left( 
\overrightarrow{v}_{12}\cdot \widehat{q}_{12}\right) ^{2}-\frac{4}{m}\Delta
_{a}\left( \widehat{q}_{12}\cdot \overrightarrow{v}_{12}\right) }\right) . 
\notag
\end{eqnarray}%
while, using the results of the previous Appendix, the latter gives two terms%
\begin{eqnarray}
\overleftrightarrow{P}^{C\left( 12\right) }\left( \overrightarrow{r}%
,t\right) &=&-\frac{m}{4}\sigma ^{D+1}\chi _{0}\left( \sigma ;n\left( 
\overrightarrow{r}\right) \right) \sum_{a}\int d\overrightarrow{v}_{1}d%
\overrightarrow{v}_{2}d\widehat{q}\;\widehat{q}\widehat{q}\left( \widehat{q}%
\cdot \overrightarrow{v}_{12}\right) \Theta \left( -\widehat{q}\cdot 
\overrightarrow{v}_{12}\right) \\
&&\times K_{a}\left( \widehat{q}\cdot \overrightarrow{v}_{12}\right) \frac{1%
}{2}\widehat{q}\cdot \left[ \left( \overrightarrow{\nabla }f_{0}\left( 
\overrightarrow{r},\overrightarrow{v}_{1};t\right) \right) f_{0}\left( 
\overrightarrow{r},\overrightarrow{v}_{2};t\right) -f_{0}\left( 
\overrightarrow{r},\overrightarrow{v}_{1};t\right) \left( \overrightarrow{%
\nabla }f_{0}\left( \overrightarrow{r},\overrightarrow{v}_{2};t\right)
\right) \right]  \notag \\
&&\times \left( -\overrightarrow{v}_{12}\cdot \widehat{q}-sgn\left( 
\overrightarrow{v}_{12}\cdot \widehat{q}\right) \sqrt{\left( \overrightarrow{%
v}_{12}\cdot \widehat{q}\right) ^{2}-\frac{4}{m}\Delta _{a}\left( \widehat{q}%
\cdot \overrightarrow{v}_{12}\right) }\right) .  \notag
\end{eqnarray}%
and%
\begin{eqnarray}
\overleftrightarrow{P}^{C\left( 13\right) }\left( \overrightarrow{r}%
,t\right) &=&\frac{m}{4}\frac{1}{2}\sigma ^{D+1}\frac{\partial \chi
_{0}\left( \sigma ;n\left( \overrightarrow{r}\right) \right) }{\partial
n\left( \overrightarrow{r}\right) }\left( \overrightarrow{\nabla }%
_{1}n\left( \overrightarrow{r}\right) \right) \cdot \sum_{a}\int d%
\overrightarrow{v}_{1}d\overrightarrow{v}_{2}d\widehat{q}\;\widehat{q}%
\widehat{q}\widehat{q}\left( \widehat{q}\cdot \overrightarrow{v}_{12}\right)
\Theta \left( -\widehat{q}\cdot \overrightarrow{v}_{12}\right) \\
&&\times K_{a}\left( \widehat{q}\cdot \overrightarrow{v}_{12}\right)
f_{0}\left( \overrightarrow{r},\overrightarrow{v}_{1};t\right) f_{0}\left( 
\overrightarrow{r},\overrightarrow{v}_{2};t\right)  \notag \\
&&\times \left( -\overrightarrow{v}_{12}\cdot \widehat{q}-sgn\left( 
\overrightarrow{v}_{12}\cdot \widehat{q}\right) \sqrt{\left( \overrightarrow{%
v}_{12}\cdot \widehat{q}\right) ^{2}-\frac{4}{m}\Delta _{a}\left( \widehat{q}%
\cdot \overrightarrow{v}_{12}\right) }\right) .  \notag
\end{eqnarray}%
so that $\overleftrightarrow{P}^{C\left( 1\right) }=\overleftrightarrow{P}%
^{C\left( 11\right) }+\overleftrightarrow{P}^{C\left( 12\right) }+%
\overleftrightarrow{P}^{C\left( 13\right) }$. However, it is seen that $%
\overleftrightarrow{P}^{C\left( 13\right) }\left( \overrightarrow{r}%
,t|f\right) =0$ because of the integral is a vector but there are no
zero-order vectors available from which to construct it. For similar
reasons, $\overleftrightarrow{P}^{C\left( 12\right) }\left( \overrightarrow{r%
},t\right) $can be simplified to%
\begin{eqnarray}
\overleftrightarrow{P}^{C\left( 12\right) }\left( \overrightarrow{r}%
,t\right) &=&-\frac{m}{8}\sigma ^{D+1}\chi _{0}\left( \sigma ;n\left( 
\overrightarrow{r}\right) \right) \left( \partial _{i}u_{j}\right)
\sum_{a}\int d\overrightarrow{v}_{1}d\overrightarrow{v}_{2}d\widehat{q}\;%
\widehat{q}\widehat{q}\left( \widehat{q}\cdot \overrightarrow{v}_{12}\right)
\Theta \left( -\widehat{q}\cdot \overrightarrow{v}_{12}\right) K_{a}\left( 
\widehat{q}\cdot \overrightarrow{v}_{12}\right) \\
&&\times \widehat{q}_{j}\frac{m}{k_{B}T}\left[ V_{1i}\frac{\partial }{%
\partial z_{1}}f_{0}\left( \overrightarrow{r},\overrightarrow{v}%
_{1};t\right) f_{0}\left( \overrightarrow{r},\overrightarrow{v}_{2};t\right)
-f_{0}\left( \overrightarrow{r},\overrightarrow{v}_{1};t\right) V_{2i}\frac{%
\partial }{\partial z_{2}}f_{0}\left( \overrightarrow{r},\overrightarrow{v}%
_{2};t\right) \right]  \notag \\
&&\times \left( -\overrightarrow{v}_{12}\cdot \widehat{q}-sgn\left( 
\overrightarrow{v}_{12}\cdot \widehat{q}\right) \sqrt{\left( \overrightarrow{%
v}_{12}\cdot \widehat{q}\right) ^{2}-\frac{4}{m}\Delta _{a}\left( \widehat{q}%
\cdot \overrightarrow{v}_{12}\right) }\right) .  \notag
\end{eqnarray}%
where $z\equiv \frac{m}{2k_{B}T}V^{2}$.

\subsection{Heat flux vector}

The expansion of the kinetic part of the heat flux vector is treated
analogous to that of the pressure tensor. It is given by%
\begin{equation}
q_{i}^{K}\left( \overrightarrow{r},t|f\right) =\frac{1}{2}m\int d%
\overrightarrow{v}\;f\left( \overrightarrow{r},\overrightarrow{v},t\right)
V_{i}V^{2},
\end{equation}%
and the zeroth order contribution vanishes by rotational symmetry.\ The
first order contribution is%
\begin{eqnarray}
q_{i}^{K(1)}\left( \overrightarrow{r},t|f\right) &=&\frac{1}{2}m\int d%
\overrightarrow{v}\;f_{1}\left( \overrightarrow{r},\overrightarrow{v}%
,t\right) V_{i}V^{2} \\
&=&\frac{1}{2}m\int d\overrightarrow{v}\;f_{0}\left( \overrightarrow{r},%
\overrightarrow{v},t\right) \left[ A^{\left( n\right) }\left( 
\overrightarrow{V}\right) V_{j}\partial _{j}n+A^{\left( T\right) }\left( 
\overrightarrow{V}\right) V_{j}\partial _{j}T\right] V_{i}V^{2}  \notag \\
&=&\frac{1}{2D}m\int d\overrightarrow{v}\;f_{0}\left( \overrightarrow{r},%
\overrightarrow{v},t\right) \left[ A^{\left( n\right) }\left( 
\overrightarrow{V}\right) V^{4}\partial _{i}n+A^{\left( T\right) }\left( 
\overrightarrow{V}\right) V^{4}\partial _{i}T\right]  \notag
\end{eqnarray}%
where the vanishing contributions have been dropped. For both the density
and temperature couplings, the important integral is%
\begin{eqnarray}
&&\frac{1}{2D}m\int d\overrightarrow{v}\;f_{0}\left( \overrightarrow{r},%
\overrightarrow{v},t\right) A^{\left( \alpha \right) }\left( \overrightarrow{%
V}\right) V^{4} \\
&=&\frac{1}{2}S_{D}\frac{1}{2D}mn\left( \frac{2k_{B}T}{m}\right)
^{2}\sum_{ij}a_{i}^{\left( \alpha \right) }c_{j}\pi ^{-D/2}\int_{0}^{\infty
}e^{-x}L_{i}^{\frac{D}{2}}\left( x\right) L_{j}^{\frac{D-2}{2}}\left(
x\right) x^{\frac{D+2}{2}}dx  \notag \\
&=&\frac{1}{2}S_{D}\frac{1}{2D}mn\left( \frac{2k_{B}T}{m}\right)
^{2}\sum_{ij}a_{i}^{\left( \alpha \right) }c_{j}\pi ^{-D/2}\int_{0}^{\infty
}e^{-x}\left( L_{i}^{\frac{D+2}{2}}\left( x\right) -L_{i-1}^{\frac{D+2}{2}%
}\left( x\right) \right) \left( L_{j}^{\frac{D+2}{2}}\left( x\right)
-2L_{j-1}^{\frac{D+2}{2}}\left( x\right) +L_{j-2}^{\frac{D+2}{2}}\left(
x\right) \right) x^{\frac{D+2}{2}}dx  \notag \\
&=&\frac{1}{2D}mn\left( \frac{2k_{B}T}{m}\right) ^{2}\sum_{k}a_{k}^{\left(
\alpha \right) }\frac{\Gamma \left( \frac{D}{2}+k+1\right) }{\Gamma \left(
k+1\right) \Gamma \left( D/2\right) }\left[ 
\begin{array}{c}
\left( \frac{1}{2}D+1+3k\right) c_{k}-2\left( \frac{1}{2}D+1+\frac{3}{2}%
k\right) c_{k+1} \\ 
+\left( \frac{1}{2}D+1+k\right) c_{k+2}-kc_{k-1}%
\end{array}%
\right]  \notag \\
&=&nk_{B}T\left( \frac{k_{B}T}{m}\right) \frac{D+2}{2}a_{1}^{\left( \alpha
\right) }\left( -1-\left( D+5\right) c_{2}+\frac{D+4}{2}c_{3}\right) +... 
\notag
\end{eqnarray}%
So%
\begin{eqnarray}
\mu ^{K} &=&\left( nk_{B}T\left( \frac{k_{B}T}{m}\right) \frac{D+2}{2}%
\right) a_{1}^{\rho }\left( 1+\left( D+5\right) c_{2}-\frac{D+4}{2}%
c_{3}\right) \\
\kappa ^{K} &=&\left( nk_{B}T\left( \frac{k_{B}T}{m}\right) \frac{D+2}{2}%
\right) a_{1}^{T}\left( 1+\left( D+5\right) c_{2}-\frac{D+4}{2}c_{3}\right) 
\notag
\end{eqnarray}%
which gives the expressions in the text if the terms $c_{2}$ and $c_{3}$ are
neglected.

The collisional part is%
\begin{eqnarray}
\overrightarrow{q}^{C}\left( \overrightarrow{r},t\right) &=&-\frac{m}{4V}%
\sigma ^{D}\sum_{a}\int d\overrightarrow{v}_{1}d\overrightarrow{v}_{2}d%
\widehat{q}\;\widehat{q}\left( \widehat{q}\cdot \overrightarrow{v}%
_{12}\right) \Theta \left( -\widehat{q}\cdot \overrightarrow{v}_{12}\right)
\\
&&K_{a}\left( \widehat{q}\cdot \overrightarrow{v}_{12}\right)
\int_{0}^{1}dy\;\chi \left( \overrightarrow{r}+\left( 1-y\right) \sigma 
\widehat{q},\overrightarrow{r}-y\sigma \widehat{q};\left[ n\right] \right)
f\left( \overrightarrow{r}+\left( 1-y\right) \sigma \widehat{q},%
\overrightarrow{v}_{1};t\right) f\left( \overrightarrow{r}-y\sigma \widehat{q%
},\overrightarrow{v}_{2};t\right)  \notag \\
&&\times \frac{1}{2}\left( \overrightarrow{V}_{1}+\overrightarrow{V}%
_{2}\right) \cdot \widehat{q}\left( -\overrightarrow{v}_{12}\cdot \widehat{q}%
-sgn\left( \overrightarrow{v}_{12}\cdot \widehat{q}\right) \sqrt{\left( 
\overrightarrow{v}_{12}\cdot \widehat{q}\right) ^{2}-\frac{4}{m}\Delta
_{a}\left( \widehat{q}\cdot \overrightarrow{v}_{12}\right) }\right) .  \notag
\end{eqnarray}%
which gives at zeroth order%
\begin{eqnarray}
\overrightarrow{q}^{C(0)}\left( \overrightarrow{r},t\right) &=&-\frac{m}{4V}%
\sigma ^{D}\chi _{0}\left( \sigma ;n\left( \overrightarrow{r}\right) \right)
\sum_{a}\int d\overrightarrow{v}_{1}d\overrightarrow{v}_{2}d\widehat{q}\;%
\widehat{q}\left( \widehat{q}\cdot \overrightarrow{v}_{12}\right) \Theta
\left( -\widehat{q}\cdot \overrightarrow{v}_{12}\right) \\
&&\times K_{a}\left( \widehat{q}\cdot \overrightarrow{v}_{12}\right)
f_{0}\left( \overrightarrow{r},\overrightarrow{v}_{1};t\right) f_{0}\left( 
\overrightarrow{r},\overrightarrow{v}_{2};t\right)  \notag \\
&&\times \frac{1}{2}\left( \overrightarrow{V}_{1}+\overrightarrow{V}%
_{2}\right) \cdot \widehat{q}\left( -\overrightarrow{v}_{12}\cdot \widehat{q}%
-sgn\left( \overrightarrow{v}_{12}\cdot \widehat{q}\right) \sqrt{\left( 
\overrightarrow{v}_{12}\cdot \widehat{q}\right) ^{2}-\frac{4}{m}\Delta
_{a}\left( \widehat{q}\cdot \overrightarrow{v}_{12}\right) }\right) .  \notag
\end{eqnarray}%
and at first order three contributions analogous to those of the pressure
tensor%
\begin{eqnarray}
\overrightarrow{q}^{C(11)}\left( \overrightarrow{r},t\right) &=&-\frac{m}{4V}%
\sigma ^{D}\chi _{0}\left( \sigma ;n\left( \overrightarrow{r}\right) \right)
\sum_{a}\int d\overrightarrow{v}_{1}d\overrightarrow{v}_{2}d\widehat{q}\;%
\widehat{q}\left( \widehat{q}\cdot \overrightarrow{v}_{12}\right) \Theta
\left( -\widehat{q}\cdot \overrightarrow{v}_{12}\right) \\
&&\times K_{a}\left( \widehat{q}\cdot \overrightarrow{v}_{12}\right) \left[
f_{0}\left( \overrightarrow{r},\overrightarrow{v}_{1};t\right) f_{1}\left( 
\overrightarrow{r},\overrightarrow{v}_{2};t\right) +f_{1}\left( 
\overrightarrow{r},\overrightarrow{v}_{1};t\right) f_{0}\left( 
\overrightarrow{r},\overrightarrow{v}_{2};t\right) \right]  \notag \\
&&\times \frac{1}{2}\left( \overrightarrow{V}_{1}+\overrightarrow{V}%
_{2}\right) \cdot \widehat{q}\left( -\overrightarrow{v}_{12}\cdot \widehat{q}%
-sgn\left( \overrightarrow{v}_{12}\cdot \widehat{q}\right) \sqrt{\left( 
\overrightarrow{v}_{12}\cdot \widehat{q}\right) ^{2}-\frac{4}{m}\Delta
_{a}\left( \widehat{q}\cdot \overrightarrow{v}_{12}\right) }\right) .  \notag
\end{eqnarray}

and%
\begin{eqnarray}
\overrightarrow{q}^{C(12)}\left( \overrightarrow{r},t\right) &=&-\frac{m}{4V}%
\sigma ^{D+1}\chi _{0}\left( \sigma ;n\left( \overrightarrow{r}\right)
\right) \sum_{a}\int d\overrightarrow{v}_{1}d\overrightarrow{v}_{2}d\widehat{%
q}\;\widehat{q}\left( \widehat{q}\cdot \overrightarrow{v}_{12}\right) \Theta
\left( -\widehat{q}\cdot \overrightarrow{v}_{12}\right)  \label{example} \\
&&\times K_{a}\left( \widehat{q}\cdot \overrightarrow{v}_{12}\right) \frac{1%
}{2}\widehat{q}\cdot \left[ \left( \overrightarrow{\nabla }f_{0}\left( 
\overrightarrow{r},\overrightarrow{v}_{1};t\right) \right) f_{0}\left( 
\overrightarrow{r},\overrightarrow{v}_{2};t\right) -f_{0}\left( 
\overrightarrow{r},\overrightarrow{v}_{1};t\right) \left( \overrightarrow{%
\nabla }f_{0}\left( \overrightarrow{r},\overrightarrow{v}_{2};t\right)
\right) \right]  \notag \\
&&\times \frac{1}{2}\left( \overrightarrow{V}_{1}+\overrightarrow{V}%
_{2}\right) \cdot \widehat{q}\left( -\overrightarrow{v}_{12}\cdot \widehat{q}%
-sgn\left( \overrightarrow{v}_{12}\cdot \widehat{q}\right) \sqrt{\left( 
\overrightarrow{v}_{12}\cdot \widehat{q}\right) ^{2}-\frac{4}{m}\Delta
_{a}\left( \widehat{q}\cdot \overrightarrow{v}_{12}\right) }\right) .  \notag
\end{eqnarray}

and%
\begin{eqnarray}
\overrightarrow{q}^{C\left( 13\right) }\left( \overrightarrow{r},t\right) &=&%
\frac{m}{4V}\frac{1}{2}\sigma ^{D+1}\frac{\partial \chi _{0}\left( \sigma
;n\left( \overrightarrow{r}\right) \right) }{\partial n\left( 
\overrightarrow{r}\right) }\left( \overrightarrow{\nabla }n\left( 
\overrightarrow{r}\right) \right) \cdot \sum_{a}\int d\overrightarrow{v}_{1}d%
\overrightarrow{v}_{2}d\widehat{q}\;\widehat{q}\widehat{q}\left( \widehat{q}%
\cdot \overrightarrow{v}_{12}\right) \Theta \left( -\widehat{q}\cdot 
\overrightarrow{v}_{12}\right) \\
&&\times K_{a}\left( \widehat{q}\cdot \overrightarrow{v}_{12}\right)
f_{0}\left( \overrightarrow{r},\overrightarrow{v}_{1};t\right) f_{0}\left( 
\overrightarrow{r},\overrightarrow{v}_{2};t\right)  \notag \\
&&\times \frac{1}{2}\left( \overrightarrow{V}_{1}+\overrightarrow{V}%
_{2}\right) \cdot \widehat{q}\left( -\overrightarrow{v}_{12}\cdot \widehat{q}%
-sgn\left( \overrightarrow{v}_{12}\cdot \widehat{q}\right) \sqrt{\left( 
\overrightarrow{v}_{12}\cdot \widehat{q}\right) ^{2}-\frac{4}{m}\Delta
_{a}\left( \widehat{q}\cdot \overrightarrow{v}_{12}\right) }\right) .  \notag
\end{eqnarray}

Now, the zeroth order heat flux vector vanishes by rotational symmetry of
the homogeneous system. Only the vector parts of $f_{1}$, i.e. those
proportional to $\overrightarrow{\nabla }n$ and $\overrightarrow{\nabla }T$,
can contribute to the first order contribution $\overrightarrow{q}^{C(11)}$.
Similarly, the second term $\overrightarrow{q}^{C(12)}$ can only depend on
those gradients by obviously the contribution proportional to $%
\overrightarrow{\nabla }n$ vanishes. Finally, the third contribution
vanishes as the integrand is odd in the total momentum.

\subsection{Heat Source}

The heat source is 
\begin{eqnarray}
\xi \left( \overrightarrow{r},t\right) &=&\frac{1}{2}\sigma
^{D-1}\sum_{a}\int d\overrightarrow{v}_{1}d\overrightarrow{v}_{2}d\widehat{q}%
\;\left( \widehat{q}\cdot \overrightarrow{v}_{12}\right) \Theta \left( -%
\widehat{q}\cdot \overrightarrow{v}_{12}\right) \\
&&\times K_{a}\left( \widehat{q}\cdot \overrightarrow{v}_{12}\right) \Delta
_{a}\left( \widehat{q}\cdot \overrightarrow{v}_{12}\right) \chi \left( 
\overrightarrow{q}_{1},\overrightarrow{q}_{2};\left[ n\right] \right)
f\left( \overrightarrow{r},\overrightarrow{v}_{1};t\right) f\left( 
\overrightarrow{r}-\sigma \widehat{q},\overrightarrow{v}_{2};t\right)  \notag
\end{eqnarray}%
so at zeroth order 
\begin{eqnarray}
\xi _{0}\left( \overrightarrow{r},t\right) &=&\frac{1}{2}\sigma ^{D-1}\chi
_{0}\left( \sigma ;n\left( \overrightarrow{r}\right) \right) \sum_{a}\int d%
\overrightarrow{v}_{1}d\overrightarrow{v}_{2}d\widehat{q}\;\left( \widehat{q}%
\cdot \overrightarrow{v}_{12}\right) \Theta \left( -\widehat{q}\cdot 
\overrightarrow{v}_{12}\right) \\
&&\times K_{a}\left( \widehat{q}\cdot \overrightarrow{v}_{12}\right) \Delta
_{a}\left( \widehat{q}\cdot \overrightarrow{v}_{12}\right) f_{0}\left( 
\overrightarrow{r},\overrightarrow{v}_{1};t\right) f_{0}\left( 
\overrightarrow{r},\overrightarrow{v}_{2};t\right) .  \notag
\end{eqnarray}%
There are, as usual, three first order contributions but as in the main
text, we separate these into $\xi _{1}\left( \overrightarrow{r},t\right)
=\xi _{0}\left[ f_{1}\right] +\xi _{11}\left( \overrightarrow{r},t\right) $
with%
\begin{eqnarray}
\xi _{0}\left[ g\right] &=&\frac{1}{2}\sigma ^{D-1}\chi _{0}\left( \sigma
;n\left( \overrightarrow{r}\right) \right) \sum_{a}\int d\overrightarrow{v}%
_{1}d\overrightarrow{v}_{2}d\widehat{q}\;\left( \widehat{q}\cdot 
\overrightarrow{v}_{12}\right) \Theta \left( -\widehat{q}\cdot 
\overrightarrow{v}_{12}\right)  \label{psi0} \\
&&\times K_{a}\left( \widehat{q}\cdot \overrightarrow{v}_{12}\right) \Delta
_{a}\left( \widehat{q}\cdot \overrightarrow{v}_{12}\right) \left[
f_{0}\left( \overrightarrow{r},\overrightarrow{v}_{1};t\right) g\left( 
\overrightarrow{r},\overrightarrow{v}_{2};t\right) +g\left( \overrightarrow{r%
},\overrightarrow{v}_{1};t\right) f_{0}\left( \overrightarrow{r},%
\overrightarrow{v}_{2};t\right) \right] ,  \notag
\end{eqnarray}%
and%
\begin{eqnarray}
\xi _{11}\left( \overrightarrow{r},t\right) &=&-\frac{1}{2}\sigma ^{D}\left(
\partial _{j}\psi _{i}\right) \cdot \sum_{a}\int d\overrightarrow{v}_{1}d%
\overrightarrow{v}_{2}d\widehat{q}\;\widehat{q}_{j}\left( \widehat{q}\cdot 
\overrightarrow{v}_{12}\right) \Theta \left( -\widehat{q}\cdot 
\overrightarrow{v}_{12}\right) \\
&&\times K_{a}\left( \widehat{q}\cdot \overrightarrow{v}_{12}\right) \Delta
_{a}\left( \widehat{q}\cdot \overrightarrow{v}_{12}\right) \chi \left( 
\overrightarrow{q}_{1},\overrightarrow{q}_{2};\left[ n\right] \right)
f_{0}\left( \overrightarrow{r},\overrightarrow{v}_{1};t\right) \frac{%
\partial }{\partial \psi _{i}}f_{0}\left( \overrightarrow{r},\overrightarrow{%
v}_{2};t\right)  \notag
\end{eqnarray}%
and note that the coefficient of $\frac{\partial \chi _{0}\left( \sigma
;n\left( \overrightarrow{r}\right) \right) }{\partial n\left( 
\overrightarrow{r}\right) }$ vanishes by spherical symmetry. The only
nonzero contribution to $\xi _{11}$comes from $\psi _{i}=\overrightarrow{u}$
and the integral must be proportional to the unit tensor so we can write $%
\xi _{11}\left( \overrightarrow{r},t\right) =\xi _{1}^{\nabla u}\left( 
\overrightarrow{r},t\right) \overrightarrow{\nabla }\cdot \overrightarrow{u}$
with%
\begin{eqnarray}
\xi _{1}^{\nabla u}\left( \overrightarrow{r},t\right) &=&-\frac{1}{2}\sigma
^{D}\sum_{a}\int d\overrightarrow{v}_{1}d\overrightarrow{v}_{2}d\widehat{q}%
\;\left( \widehat{q}\cdot \overrightarrow{v}_{12}\right) \Theta \left( -%
\widehat{q}\cdot \overrightarrow{v}_{12}\right) \\
&&\times K_{a}\left( \widehat{q}\cdot \overrightarrow{v}_{12}\right) \Delta
_{a}\left( \widehat{q}\cdot \overrightarrow{v}_{12}\right) \chi \left( 
\overrightarrow{q}_{1},\overrightarrow{q}_{2};\left[ n\right] \right)
f_{0}\left( \overrightarrow{r},\overrightarrow{v}_{1};t\right) \widehat{q}%
_{j}\frac{\partial }{\partial u_{j}}f_{0}\left( \overrightarrow{r},%
\overrightarrow{v}_{2};t\right) .  \notag
\end{eqnarray}

\bigskip

\section{Generating function for the HCS}

\label{AppGeneratingFunction}

The definition of the couplings is%
\begin{eqnarray}
I_{rs,k} &=&-n^{-1}A_{k}^{-1}\int d\overrightarrow{v}_{1}d2\;L_{k}^{\frac{D-2%
}{2}}\left( \frac{m}{2k_{B}T}v_{1}^{2}\right) \overline{T}_{-}\left(
12\right) \left( \frac{2k_{B}T}{m}\right) ^{-D}f_{M}\left( \overrightarrow{v}%
_{1}\right)  \label{g0} \\
&&\times f_{M}\left( \overrightarrow{v}_{2}\right) L_{r}^{\frac{D-2}{2}%
}\left( \frac{m}{2k_{B}T}v_{1}^{2}\right) L_{s}^{\frac{D-2}{2}}\left( \frac{m%
}{2k_{B}T}v_{1}^{2}\right) .  \notag
\end{eqnarray}%
which is equivalent to 
\begin{eqnarray}
I_{rs,k} &=&n^{-1}A_{k}^{-1}\int d\overrightarrow{v}_{1}d2\;\left( \frac{%
2k_{B}T}{m}\right) ^{-D}f_{M}\left( \overrightarrow{v}_{1}\right)
f_{M}\left( \overrightarrow{v}_{2}\right) \\
&&\times L_{r}^{\frac{D-2}{2}}\left( \frac{m}{2k_{B}T}v_{1}^{2}\right)
L_{s}^{\frac{D-2}{2}}\left( \frac{m}{2k_{B}T}v_{1}^{2}\right) T_{+}\left(
12\right) L_{k}^{\frac{D-2}{2}}\left( \frac{m}{2k_{B}T}v_{1}^{2}\right) . 
\notag
\end{eqnarray}

The associated Laguerre polynomials are generated by%
\begin{equation}
L_{n}^{\alpha }\left( x\right) =\frac{1}{n!}\lim_{z\rightarrow 0}\frac{%
\partial ^{n}}{\partial z^{n}}\frac{1}{\left( 1-z\right) ^{\alpha +1}}\exp
\left( -\frac{xz}{1-z}\right) .  \label{g1}
\end{equation}%
so that we have%
\begin{equation}
I_{rs,k}=-nA_{k}^{-1}\left( \frac{m}{2k_{B}T}\right) ^{-1/2}\sigma ^{D-1}%
\frac{1}{r!s!k!}\lim_{z_{1}\rightarrow 0}\lim_{z_{2}\rightarrow
0}\lim_{x\rightarrow 0}\frac{\partial ^{r}}{\partial z_{1}^{r}}\frac{%
\partial ^{s}}{\partial z_{2}^{s}}\frac{\partial ^{k}}{\partial x^{k}}\left[
\sum_{a}G_{a}\left( \Delta _{a}\right) -G_{0}\right]  \notag
\end{equation}%
with%
\begin{eqnarray}
G_{0} &=&\frac{1}{\left( 1-z_{1}\right) ^{\frac{D}{2}}}\frac{1}{\left(
1-z_{2}\right) ^{\frac{D}{2}}}\frac{1}{\left( 1-x\right) ^{\frac{D}{2}}}\pi
^{-D}  \label{g4} \\
&&\times \int d\overrightarrow{v}_{1}d\overrightarrow{v}_{2}d\widehat{q}%
\;\exp \left( -\frac{1}{1-z_{1}}v_{1}^{2}-\frac{1}{1-z_{2}}v_{2}^{2}\right)
\left( \widehat{q}\cdot \overrightarrow{v}_{12}\right) \Theta \left( -%
\widehat{q}\cdot \overrightarrow{v}_{12}\right) \exp \left( -\frac{x}{1-x}%
v_{1}^{2}\right)  \notag \\
G_{a}\left( \Delta \right) &=&\frac{1}{\left( 1-z_{1}\right) ^{\frac{D}{2}}}%
\frac{1}{\left( 1-z_{2}\right) ^{\frac{D}{2}}}\frac{1}{\left( 1-x\right) ^{%
\frac{D}{2}}}\pi ^{-D}\int d\overrightarrow{v}_{1}d\overrightarrow{v}_{2}d%
\widehat{q}\;K_{a}^{\ast }\left( \widehat{q}\cdot \overrightarrow{v}%
_{12}\right)  \notag \\
&&\times \exp \left( -\frac{1}{1-z_{1}}v_{1}^{2}-\frac{1}{1-z_{2}}%
v_{2}^{2}\right) \left( \widehat{q}\cdot \overrightarrow{v}_{12}\right)
\Theta \left( -\widehat{q}\cdot \overrightarrow{v}_{12}\right) \widehat{b}%
_{a}\exp \left( -\frac{x}{1-x}v_{1}^{2}\right) .  \notag
\end{eqnarray}%
To evaluate the second quantity, note that%
\begin{equation}
v_{1}^{\prime 2}=V^{2}+\overrightarrow{V}\cdot \overrightarrow{v}-%
\overrightarrow{V}\cdot \widehat{q}\left( \overrightarrow{v}\cdot \widehat{q}%
+sgn\left( \overrightarrow{v}\cdot \widehat{q}\right) \sqrt{\left( 
\overrightarrow{v}\cdot \widehat{q}\right) ^{2}-2\beta \Delta \left( 
\overrightarrow{v}\cdot \widehat{q}\right) }\right) +\frac{1}{4}v^{2}-\frac{%
\beta }{2}\Delta \left( \overrightarrow{v}\cdot \widehat{q}\right) .
\label{g5}
\end{equation}%
A tedious calculation to complete the square in $V$ gives%
\begin{eqnarray}
G_{a} &=&\frac{1}{\left( 1-z_{1}\right) ^{\frac{D}{2}}}\frac{1}{\left(
1-z_{2}\right) ^{\frac{D}{2}}}\frac{1}{\left( 1-x\right) ^{\frac{D}{2}}}\pi
^{-D}\int d\overrightarrow{V}d\overrightarrow{v}d\widehat{q}\;\left( 
\widehat{q}\cdot \overrightarrow{v}\right) \Theta \left( -\widehat{q}\cdot 
\overrightarrow{v}\right)  \label{g9} \\
&&\times K_{a}^{\ast }\left( \widehat{q}\cdot \overrightarrow{v}_{12}\right)
\exp \left( \frac{\left( 2-z_{2}-z_{1}\right) x}{2-x-z_{2}-z_{1}+xz_{1}z_{2}}%
\frac{1}{2}\Delta _{a}^{\ast }\left( \overrightarrow{v}\cdot \widehat{q}%
\right) \right)  \notag \\
&&\times \exp \left( -\left( \frac{1}{1-z_{1}}+\frac{1}{1-z_{2}}+\frac{x}{1-x%
}\right) V^{2}\right) \exp \left( -\frac{1-z_{1}x}{%
2-x-z_{2}-z_{1}+xz_{1}z_{2}}v^{2}\right)  \notag \\
&&\times \exp \left( \frac{1}{2}\frac{\left( z_{2}-z_{1}\right) x}{%
2-x-z_{2}-z_{1}+xz_{1}z_{2}}\left[ \allowbreak \allowbreak \left( 
\overrightarrow{v}\cdot \widehat{q}\right) ^{2}+\allowbreak \allowbreak 
\overrightarrow{v}\cdot \widehat{q}sgn\left( \overrightarrow{v}\cdot 
\widehat{q}\right) \sqrt{\left( \overrightarrow{v}\cdot \widehat{q}\right)
^{2}-2\Delta _{a}^{\ast }\left( \overrightarrow{v}\cdot \widehat{q}\right) }%
\right] \right)  \notag
\end{eqnarray}%
Performing the $V$ integration and the $D-1$ $v$ integrations in directions
perpendicular to $\widehat{q}$ and finally the $\widehat{q}$ integral leave
the final result%
\begin{eqnarray}
G_{a} &=&-\frac{1}{2}\pi ^{-1/2}S_{D}\left( 1-z_{1}x\right) ^{-\frac{1}{2}%
D}\allowbreak \left( \frac{1-z_{1}x}{2-x-z_{2}-z_{1}+xz_{1}z_{2}}\right) ^{%
\frac{1}{2}}  \label{g10a} \\
&&\times \int_{0}^{\infty }du\;K_{a}^{\ast }\left( \sqrt{u}\right) \exp
\left( \frac{\left( 2-z_{2}-z_{1}\right) x}{2-x-z_{2}-z_{1}+xz_{1}z_{2}}%
\frac{1}{2}\Delta _{a}^{\ast }\left( \sqrt{u}\right) \right)  \notag \\
&&\times \exp \left( -\frac{1}{2}\left( \frac{2-z_{2}x-z_{1}x}{%
2-x-z_{2}-z_{1}+xz_{1}z_{2}}u\right) \right)  \notag \\
&&\times \exp \left( \frac{1}{2}\frac{\left( z_{2}-z_{1}\right) x}{%
2-x-z_{2}-z_{1}+xz_{1}z_{2}}\sqrt{u}\sqrt{u-2\Delta _{a}^{\ast }\left( \sqrt{%
u}\right) }\right) .  \notag
\end{eqnarray}

We will also need the generating function for the case that no collision
occurs 
\begin{eqnarray}
G_{0} &=&\frac{1}{\left( 1-z_{1}\right) ^{\frac{D}{2}}}\frac{1}{\left(
1-z_{2}\right) ^{\frac{D}{2}}}\frac{1}{\left( 1-x\right) ^{\frac{D}{2}}}\pi
^{-D}  \label{g11} \\
&&\times \int d\overrightarrow{v}_{1}d\overrightarrow{v}_{2}d\widehat{q}%
\;\exp \left( -\frac{1}{1-z_{1}}v_{1}^{2}-\frac{1}{1-z_{2}}v_{2}^{2}\right) 
\widehat{q}\cdot \overrightarrow{v}_{12}\Theta \left( -\widehat{q}\cdot 
\overrightarrow{v}_{12}\right) \exp \left( -\frac{x}{1-x}v_{1}^{2}\right) 
\notag
\end{eqnarray}%
Completing the square and performing the simple Gaussian integrals gives%
\begin{equation}
G_{0}=-\frac{1}{2}\pi ^{-1/2}S_{D}\left( 1-z_{1}x\right) ^{-\frac{D+1}{2}%
}\left( 2-x-z_{2}-z_{1}+xz_{1}z_{2}\right) ^{\frac{1}{2}}.
\end{equation}

\section{Evaluation of integrals}

\label{AppEvaluation}

To illustrate the method used to evaluate the many integrals required in
this work, consider the quantity $\overrightarrow{q}^{C(12)}\left( 
\overrightarrow{r},t\right) $ defined in eq.(\ref{example}) and for
convenience repeated here 
\begin{eqnarray}
\overrightarrow{q}^{C(12)}\left( \overrightarrow{r},t\right) &=&-\frac{m}{4}%
\sigma ^{D+1}\chi _{0}\left( \sigma ;n\left( \overrightarrow{r}\right)
\right) \sum_{a}\int d\overrightarrow{v}_{1}d\overrightarrow{v}_{2}d\widehat{%
q}\;\widehat{q}\left( \widehat{q}\cdot \overrightarrow{v}_{12}\right) \Theta
\left( -\widehat{q}\cdot \overrightarrow{v}_{12}\right) \\
&&\times K_{a}\left( \widehat{q}\cdot \overrightarrow{v}_{12}\right) \frac{1%
}{2}\widehat{q}\cdot \left[ \left( \overrightarrow{\nabla }f_{0}\left( 
\overrightarrow{r},\overrightarrow{v}_{1};t\right) \right) f_{0}\left( 
\overrightarrow{r},\overrightarrow{v}_{2};t\right) -f_{0}\left( 
\overrightarrow{r},\overrightarrow{v}_{1};t\right) \left( \overrightarrow{%
\nabla }f_{0}\left( \overrightarrow{r},\overrightarrow{v}_{2};t\right)
\right) \right]  \notag \\
&&\times \frac{1}{2}\left( \overrightarrow{V}_{1}+\overrightarrow{V}%
_{2}\right) \cdot \widehat{q}\left( -\overrightarrow{v}_{12}\cdot \widehat{q}%
-sgn\left( \overrightarrow{v}_{12}\cdot \widehat{q}\right) \sqrt{\left( 
\overrightarrow{v}_{12}\cdot \widehat{q}\right) ^{2}-\frac{4}{m}\Delta
_{a}\left( \widehat{q}\cdot \overrightarrow{v}_{12}\right) }\right) .  \notag
\end{eqnarray}%
As stated previously, the only nonzero contribution comes through the
temperature so we replace 
\begin{equation}
\overrightarrow{\nabla }f_{0}\rightarrow \left( \overrightarrow{\nabla }%
T\right) \frac{\partial }{\partial T}f_{0}.
\end{equation}%
Keeping terms up to linear order in $c_{2}$ and defining the quantities 
\begin{eqnarray}
z_{i} &=&\frac{m}{2k_{B}T}V_{i}^{2} \\
\overrightarrow{V} &=&\sqrt{\frac{m}{2k_{B}T}}\left( \overrightarrow{v}_{1}+%
\overrightarrow{v}_{2}\right)  \notag \\
\overrightarrow{v} &=&\sqrt{\frac{m}{2k_{B}T}}\left( \overrightarrow{v}_{1}-%
\overrightarrow{v}_{2}\right)  \notag
\end{eqnarray}%
we find that%
\begin{eqnarray}
&&\left( \frac{\partial }{\partial T}f^{(0)}\left( \overrightarrow{r},%
\overrightarrow{v}_{1};t\right) \right) f^{(0)}\left( \overrightarrow{r},%
\overrightarrow{v}_{2};t\right) -f^{(0)}\left( \overrightarrow{r},%
\overrightarrow{v}_{1};t\right) \frac{\partial }{\partial T}f^{(0)}\left( 
\overrightarrow{r},\overrightarrow{v}_{2};t\right) \\
&=&-\frac{1}{T}\left[ \left( z_{1}\frac{\partial }{\partial z_{1}}%
f^{(0)}\left( \overrightarrow{r},\overrightarrow{v}_{1};t\right) \right)
f^{(0)}\left( \overrightarrow{r},\overrightarrow{v}_{2};t\right)
-f^{(0)}\left( \overrightarrow{r},\overrightarrow{v}_{1};t\right) z_{2}\frac{%
\partial }{\partial z_{2}}f^{(0)}\left( \overrightarrow{r},\overrightarrow{v}%
_{2};t\right) \right]  \notag \\
&=&-\frac{1}{T}\left[ 
\begin{array}{c}
\left( -z_{1}+z_{2}\right) \left( 1+c_{2}\left[ 
\begin{array}{c}
\left( \frac{1}{4}D^{2}+\frac{1}{2}D\right) -\left( D+2\right) \left( V^{2}+%
\frac{1}{4}v^{2}\right) \\ 
+\left( V^{4}+\frac{1}{2}V^{2}v^{2}+\frac{1}{16}v^{4}+\left( \overrightarrow{%
V}\cdot \overrightarrow{v}\right) ^{2}\right)%
\end{array}%
\right] \right) \\ 
+z_{1}c_{2}\left( -\frac{1}{2}\left( D+2\right) +z_{1}\right)
-z_{2}c_{2}\left( -\frac{1}{2}\left( D+2\right) +z_{2}\right)%
\end{array}%
\right] f_{M}\left( \overrightarrow{r},\overrightarrow{v}_{1};t\right)
f_{M}\left( \overrightarrow{r},\overrightarrow{v}_{2};t\right)  \notag \\
&=&\frac{1}{T}2\overrightarrow{V}\cdot \overrightarrow{v}\left( 1+c_{2}\left[
\begin{array}{c}
\frac{1}{4}\left( D+2\right) ^{2}-\left( D+4\right) \left( V^{2}+\frac{1}{4}%
v^{2}\right) \\ 
+\left( V^{4}+\frac{1}{2}V^{2}v^{2}+\frac{1}{16}v^{4}+\left( \overrightarrow{%
V}\cdot \overrightarrow{v}\right) ^{2}\right)%
\end{array}%
\right] \right) f_{M}\left( \overrightarrow{r},\overrightarrow{v}%
_{1};t\right) f_{M}\left( \overrightarrow{r},\overrightarrow{v}_{2};t\right)
\notag
\end{eqnarray}%
Substituting into the original expression and changing integration variables
gives 
\begin{eqnarray}
q_{i}^{C(12)}\left( \overrightarrow{r},t\right) &=&-\frac{m}{4}n^{2}\sigma
^{D+1}\chi _{0}\left( \sigma ;n\left( \overrightarrow{r}\right) \right)
\left( \partial _{j}T\right) \left( \frac{2k_{B}T}{m}\right) ^{3/2}\pi
^{-D}\sum_{a}\int d\overrightarrow{V}dvd\widehat{q}\;\widehat{q}_{i}\widehat{%
q}_{j}\left( \widehat{q}\cdot \overrightarrow{v}\right) \Theta \left( -%
\widehat{q}\cdot \overrightarrow{v}\right) \\
&&\times K_{a}\left( \sqrt{\frac{2k_{B}T}{m}}\widehat{q}\cdot 
\overrightarrow{v}\right) \frac{1}{2}\frac{1}{T}2\overrightarrow{V}\cdot 
\overrightarrow{v}\left( 1+c_{2}\left[ 
\begin{array}{c}
\frac{1}{4}\left( D+2\right) ^{2}-\left( D+4\right) \left( V^{2}+\frac{1}{4}%
v^{2}\right) \\ 
+\left( V^{4}+\frac{1}{2}V^{2}v^{2}+\frac{1}{16}v^{4}+\left( \overrightarrow{%
V}\cdot \overrightarrow{v}\right) ^{2}\right)%
\end{array}%
\right] \right) e^{-2V^{2}-\frac{1}{2}v^{2}}  \notag \\
&&\times \frac{1}{2}\left( \overrightarrow{V}\cdot \widehat{q}\right) \left(
-\overrightarrow{v}\cdot \widehat{q}-sgn\left( \overrightarrow{v}\cdot 
\widehat{q}\right) \sqrt{\left( \overrightarrow{v}\cdot \widehat{q}\right)
^{2}-2\Delta _{a}^{\ast }\left( \widehat{q}\cdot \overrightarrow{v}\right) }%
\right) .  \notag
\end{eqnarray}%
Taking $\widehat{v}$ to be the $x$-direction for the $V$ integral, this
becomes 
\begin{eqnarray}
q_{i}^{C(12)}\left( \overrightarrow{r},t\right) &=&-\frac{m}{8}n^{2}\sigma
^{D+1}\chi _{0}\left( \sigma ;n\left( \overrightarrow{r}\right) \right)
\left( \partial _{j}T\right) \left( \frac{2k_{B}T}{m}\right) ^{3/2}\frac{1}{T%
}\pi ^{-D}\sum_{a}\int d\overrightarrow{V}d\overrightarrow{v}d\widehat{q}\;%
\widehat{q}_{i}\widehat{q}_{j}\left( \widehat{q}\cdot \overrightarrow{v}%
\right) ^{2}\Theta \left( -\widehat{q}\cdot \overrightarrow{v}\right) \\
&&\times K_{a}^{\ast }\left( \widehat{q}\cdot \overrightarrow{v}\right)
V_{x}^{2}\left( 1+c_{2}\left[ 
\begin{array}{c}
\frac{1}{4}\left( D+2\right) ^{2}-\left( D+4\right) \left( V^{2}+\frac{1}{4}%
v^{2}\right) \\ 
+\left( V^{4}+\frac{1}{2}V^{2}v^{2}+\frac{1}{16}v^{4}+V_{x}^{2}v^{2}\right)%
\end{array}%
\right] \right) e^{-2V^{2}-\frac{1}{2}v^{2}}  \notag \\
&&\times \left( -\overrightarrow{v}\cdot \widehat{q}-sgn\left( 
\overrightarrow{v}\cdot \widehat{q}\right) \sqrt{\left( \overrightarrow{v}%
\cdot \widehat{q}\right) ^{2}-2\Delta _{a}^{\ast }\left( \widehat{q}\cdot 
\overrightarrow{v}\right) }\right) .  \notag
\end{eqnarray}%
The method now is to take $\widehat{q}$ to be the $x$-direction in the $v$
integrals and to make the following substitutions:%
\begin{eqnarray}
V^{2} &\rightarrow &V_{x}^{2}+V_{\bot }^{2} \\
v^{2} &\rightarrow &v_{x}^{2}+v_{\bot }^{2}  \notag
\end{eqnarray}%
and to integrate over $V_{x}$, $V_{\bot }$and $v_{\bot }$. These integrals
are performed by using symbolic manipulation to expand the kernal and to
make make the replacement%
\begin{equation}
\pi ^{-D}V_{x}^{l}V_{\bot }^{m}v_{\bot }^{n}\rightarrow \frac{1}{\sqrt{2\pi }%
}\left( 2^{-\frac{l+1}{2}}\sqrt{\frac{2}{\pi }}\Gamma \left( \frac{l+1}{2}%
\right) \right) \left( 2^{-\frac{m}{2}}\frac{\Gamma \left( \frac{D-1+m}{2}%
\right) }{\Gamma \left( \frac{D-1}{2}\right) }\right) \left( 2^{\frac{n}{2}}%
\frac{\Gamma \left( \frac{D-1+n}{2}\right) }{\Gamma \left( \frac{D-1}{2}%
\right) }\right)
\end{equation}%
to get%
\begin{eqnarray}
q_{i}^{C(12)}\left( \overrightarrow{r},t\right) &=&-\frac{m}{4}n^{2}\sigma
^{D+1}\chi _{0}\left( \partial _{j}T\right) \left( \frac{k_{B}T}{m}\right)
^{3/2}\frac{1}{T\sqrt{\pi }}\sum_{a}\int_{-\infty }^{0}dv_{x}d\widehat{q}\;%
\widehat{q}_{i}\widehat{q}_{j}v_{x}^{2} \\
&&\times K_{a}^{\ast }\left( v_{x}\right) \left( \frac{1}{4}+\frac{1}{64}%
c_{2}\left( v_{x}^{4}-2v_{x}^{2}-9\right) \right) e^{-\frac{1}{2}%
v_{x}^{2}}\left( -v_{x}-sgn\left( v_{x}\right) \sqrt{v_{x}^{2}-2\Delta
_{a}^{\ast }\left( v_{x}\right) }\right)  \notag \\
&=&-\frac{m}{4V}n^{2}\sigma ^{D-1}\chi _{0}\left( \partial _{i}T\right)
\left( \frac{k_{B}T}{m}\right) ^{3/2}\frac{S_{D}}{D\sqrt{\pi }}  \notag \\
&&\times \sum_{a}\int_{0}^{\infty }K_{a}\left( v_{x}\right) v_{x}^{2}e^{-%
\frac{1}{2}v_{x}^{2}}\left( \frac{1}{4}+\frac{1}{64}c_{2}\left(
v_{x}^{4}-2v_{x}^{2}-9\right) \right) \left( \sqrt{v_{x}^{2}-2\Delta
_{a}^{\ast }\left( v_{x}\right) }+v_{x}\right) dv_{x}  \notag
\end{eqnarray}%
We can isolate the contribution due to $\Delta _{a}^{\ast }\left(
v_{x}\right) \neq 0$ by subtracting the $\Delta _{a}^{\ast }\left(
v_{x}\right) =0$ term and using $\sum_{a}K_{a}^{\ast }\left( v_{x}\right) =1$
to get%
\begin{eqnarray}
q_{i}^{C(12)}\left( \overrightarrow{r},t\right) &=&-\frac{m}{4}n^{2}\sigma
^{D+1}\chi _{0}\left( \partial _{i}T\right) \left( \frac{k_{B}T}{m}\right)
^{3/2}\frac{S_{D}}{DT\sqrt{\pi }} \\
&&\times \left[ 1+\frac{7}{16}c_{2}+\frac{1}{4}\sum_{a}\int_{0}^{\infty
}K_{a}^{\ast }\left( v_{x}\right) v_{x}^{2}e^{-\frac{1}{2}v_{x}^{2}}\left( 1+%
\frac{1}{16}c_{2}\left( v_{x}^{4}-2v_{x}^{2}-9\right) \right) \left( \sqrt{%
v_{x}^{2}-2\Delta _{a}^{\ast }\left( v_{x}\right) }-v_{x}\right) dv_{x}%
\right]  \notag
\end{eqnarray}

The only further complication is that some integrals involve kernals of the
form $h(V^{2})\left( \overrightarrow{V}\cdot \widehat{q}\right) ^{2}\left( 
\overrightarrow{V}\cdot \overrightarrow{v}\right) ^{2}$. These can be
handled by using the substitution%
\begin{equation}
h(V^{2})V_{i}V_{j}V_{k}V_{l}\rightarrow h(V^{2})V^{4}\frac{1}{D^{2}+2D}%
\left( \delta _{ij}\delta _{lm}+\delta _{il}\delta _{jm}+\delta _{im}\delta
_{jl}\right) .
\end{equation}

\bigskip

\bibliographystyle{prsty}
\bibliography{physics}

\end{document}